\documentstyle[aps,preprint,psfig]{revtex}

\title
{Mixed Quantum-Classical versus Full Quantum Dynamics: Coupled
Quasiparticle-Oscillator System}

\author{Holger Schanz\footnote{e-mail: schanz@physik.hu-berlin.de} and Bernd Esser} 

\address{Institut f\"ur Physik,
Humboldt-Universit\"at,
Invalidenstr.\ 110,
10 115 Berlin, Germany}

\date{September 25, 1996}

\begin{document}
\thispagestyle{empty}
\maketitle

\begin{abstract}
  The relation between the dynamical properties of a coupled
  quasiparticle-oscillator system in the mixed quantum-classical and
  fully quantized descriptions is investigated. The system is
  considered to serve as a model system for applying a stepwise
  quantization. Features of the nonlinear dynamics of the mixed
  description such as the presence of a separatrix structure or
  regular and chaotic motion are shown to be reflected in the
  evolution of the quantum state vector of the fully quantized system.
  In particular it is demonstrated how wave packets propagate along
  the separatrix structure of the mixed description and that chaotic
  dynamics leads to a strongly entangled quantum state vector. Special
  emphasis is given to view the system from a dynamical
  Born-Oppenheimer approximation defining integrable reference
  oscillators and elucidating the role of the nonadiabatic couplings
  which complements this approximation into a rigorous quantization
  scheme.
\end{abstract}

\section{Introduction}
The relation between the quantum and classical dynamics of nonlinear
systems includes a specific side in the correspondence between the
dynamical properties of systems treated in the mixed and fully
quantized descriptions. Various aspects of the correspondences between
classical nonlinear systems on the one side and their fully quantized
counterparts on the other have been intensively investigated in the
last decade (see e.\ g.\ \cite{Hel91,BTU,Rei}). In many systems
relevant for molecular and condensed matter physics the direct
quantization of the full system in one step is, however, not possible
from a practical point of view.  As a rule such systems divide
naturally into interacting subsystems.  Then a stepwise quantization
is applied resulting in a mixed description, in which one of the
subsystems is treated in the quantum and the other in the classical
context.  Furthermore in complex systems the mixed description is
often necessary for understanding global dynamical properties, e.\ g.
the presence of bifurcations and separatrix structures dividing the
solution manifold into characteristic parts, before for a selected
energy region the full quantization can be performed.

This stepwise quantization is the basic idea on which the Born -
Oppenheimer approximation developed in the early days of quantum
mechanics for the quantization of systems dividing into subsystems is
based. As is well known this approximation can be complemented into a
rigorous scheme, if the nonadiabatic couplings are included \cite{BH}.
These couplings can be the source of nonintegrability and chaos of
systems treated in the mixed quantum-classical description
\cite{BK,BE}.  Then the problem of the quantum-classical
correspondences arises on the level of the relation between the
dynamical properties of the mixed and fully quantized descriptions
\cite{BE}. 

In this paper we consider this dynamic relation for the particular
model of a quasiparticle moving between two sites and coupled to
oscillators. This is an important model system with applications such
as excitons moving in molecular aggregates and coupled to vibrations,
see e.~g.\  \cite{W}. It has also attracted widespread attention in the
context of the spin-boson Hamiltonian and its classical-quantum
phase space behavior and correspondence (see e. g. \cite{GH,K,Sto} and
references therein).  Hence it seems appropriate to use this system as
a model to analyze the relation between the mixed and fully quantized
descriptions. Treating the oscillators in the classical or quantum
contexts, whereas the quasiparticle moving between two sites is a
quantum object from the beginning, one arrives at mixed and fully
quantized levels of description.  We have recently investigated the
dynamical properties of this model in the mixed description by
integrating the corresponding Bloch-oscillator equations and
demonstrated the presence of a phase space with an underlying
separatrix structure for overcritical coupling and chaos developing
from the region of the hyperbolic point at the center of this
structure. For increasing total energy chaos spreads over the product
phase space of the system constituted by the Bloch sphere and
oscillator plane, leaving only regular islands in the region of the
antibonding states \cite{ES}.  Here we consider the problem of the
relation between the dynamics in the mixed and fully quantum levels of
description of the coupled quasiparticle-oscillator motion.
Investigating this relation we focus on the adiabatic parameter range,
where the closest correspondence between the classical and quantum
aspects of the oscillator dynamics can be expected.  Although several
aspects of the dynamics of the system have been considered
\cite{GH,Sto}, there exists no systematic investigation in the
adiabatic parameter range.  In particular, such an investigation
requires the numerical determination of a large number of eigenstates
for the fully quantized system.  The stationary properties of these
states were reported in \cite{inc}. In this paper these states are
used to compute the dynamics of the fully quantized system and to
compare the quantum evolution with the dynamics of the mixed
description where the oscillator is treated classically. Performing
this comparison we use both the fixed and adiabatic basis sets in the
mixed description. The latter basis set is of particular importance to
clarify the role of the nonadiabatic couplings in the formation of the
dynamics. We demonstrate the effect of the separatrix structure of the
mixed description in the oscillator wave packet propagation of the
fully quantized version, of dynamical subsystem correlations deriving
from the separatrix structure and how the chaotic phase space regions
of the system in the mixed description show up in the nonstationary
properties of the time dependent full quantum state vector.

In section \ref{model} the model will be specified in detail. The
mixed quantum-classical description is discussed in section \ref{mqcd}
including the derivation of the equations of motion, the fixed point
structure and the dynamical properties of the system on this level of
description.  In section \ref{qevol} the evolution of the fully
quantized system is presented and compared to the dynamics in the
mixed description.

\section{The Model}\label{model}
We consider a quasiparticle coupled to oscillator degrees of
freedom. The quasiparticle is specified as a molecular exciton in a tight 
binding representation and can be substituted by any other quantum object 
moving between discrete sites and described by a tight binding Hamiltonian 
of the same structure. The system has the Hamiltonian   
\begin{equation}\label{htot}
H^{{\rm (tot)}} = H^{{\rm (exc)}} + H^{{\rm (vib)}} + H^{{\rm (int)}},
\end{equation}
where $H^{{\rm (exc)}}$, $H^{{\rm (vib)}}$ and $H^{{\rm (int)}}$ are the excitonic, vibronic and interaction
parts, respectively. $H^{{\rm (exc)}}$ represents the quantum subsystem, which is taken
in the site representation 
\begin{equation}\label{hex}
H^{{\rm (exc)}} = \sum_n \epsilon_n |c_n|^2 + \sum_{n \ne m} V_{nm} c_n^*\,c_m,
\end{equation}
where $c_n$ is the quantum probability amplitude of the exciton to
occupy the $n$-th molecule and $V_{nm}$ the transfer matrix element.
For the intramolecular vibrations coupling to the exciton we use the
harmonic approximation in $H^{{\rm (vib)}}$
\begin{equation}\label{hvi}
H^{{\rm (vib)}} = {1 \over 2} \sum_n (p_n^2 + \omega_n^2 q_n^2).
\end{equation}
Here  $q_n$, $p_n$ and $\omega_n$ are the coordinate, the canonic conjugate momentum
and frequency of the intramolecular vibration of the $n$-th molecule, 
respectively. The interaction Hamiltonian $H^{{\rm (int)}}$ represents the dependence 
of the exciton energy on the intramolecular configuration for which we use 
the first order expansion in $q_n$    
\begin{equation}\label{int}
H^{{\rm (int)}} = \sum_{n} \gamma_n q_n |c_n|^2,
\end{equation}
where $\gamma_n$ are the coupling constants. The interaction is restricted to a 
single oscillator at each molecule.
The case of a symmetric two site system $n=1,2$, e.~g.\  an exciton in a molecular 
dimer constituted by two identical monomers, is considered in what follows. 
For this case we set $\epsilon_1=\epsilon_2$, $\omega_1=\omega_2$, $\gamma_1=\gamma_2$ and 
$V_{12}=V_{21}=-V$, $V>0$. Then by introducing for the vibronic subsystem 
the new coordinates and momenta
\begin{equation}\label{qp}
q_\pm := {q_2 \pm q_1 \over \sqrt{2} }\hspace*{1cm}
p_\pm := {p_2 \pm p_1 \over \sqrt{2} },
\end{equation}
and for the excitonic subsystem the Bloch variables
\begin{equation}\label{bl}
x=\rho_{21}+\rho_{12},\hspace*{1cm}
y=i(\rho_{21}-\rho_{12}),\hspace*{1cm}
z=\rho_{22}-\rho_{11},
\end{equation}
where $\rho_{mn}$ is the density matrix of the excitonic subsystem 
\begin{equation}\label{dm}
\rho_{mn}=c_n^*c_m,
\end{equation}
the relevant part of (\ref{htot})-(\ref{int}) connected with the exciton 
coupled to the $q_-$ vibration is obtained in the form
\begin{equation}\label{hsymx}
H_- = -Vx + {1 \over 2} (p_-^2 + \omega^2 q_-^2) +
{\gamma q_- z \over  \sqrt{2}}\,.
\end{equation} 
The part corresponding to $q_+$ is not coupled to the exciton and omitted. 

The Hamiltonian (\ref{hsymx}) can be represented as an operator in the
space of the two dimensional vectors $C=(c_1,c_2)$ constituted by the
excitonic amplitudes $c_n$ by using the standard Pauli spin matrices
$\sigma_i$ $(i=x,y,z)$. Passing in (\ref{hsymx}) to dimensionless
variables by measuring $H$ in units of $2V$ and replacing $q_-$, $p_-$
by
\begin{equation}\label{gqp}
Q := \sqrt{2V} q_- \hspace*{1cm} P := {1 \over \sqrt{2V}} p_-, 
\end{equation}
one finally obtains
\begin{equation}\label{hsym}
H = -{\sigma_x\over2} + {1 \over 2} (P^2 + r^2 Q^2) +
\sqrt{p\over 2}r\,Q\sigma_z\, .
\end{equation}
Here 
\begin{equation}\label{par}
p = {\gamma^2\over 2V\omega^2}
\end{equation}
represents the dimensionless excitonic-vibronic coupling
and 
\begin{equation}\label{rad}
r = {\omega \over 2V} 
\end{equation}
is the adiabatic parameter measuring the relative strength of quantum
effects in both subsystems.  We focus on the adiabatic case $r\ll 1$
when the vibronic subsystem can be described in the classical
approximation. To make contact with the dynamical features following
from the adiabatic approximation we derive the basic equations in both
the fixed and the adiabatic bases.

\section{Mixed Quantum-Classical Description}\label{mqcd}
\subsection{Fixed Basis} 
In this case the basis states are given by the fixed molecule sites
$|n\rangle$ . Representing the excitonic state by $|\psi\rangle = \sum_n c_n
|n\rangle$, inserting it into the time dependent Schroedinger equation for
(\ref{hsym}) and using (\ref{bl}) the quantum equations of motion for
the excitonic subsystem describing the transfer dynamics between the
two sites are obtained.  The classical equations for the dynamics of
the oscillator are found by passing to the expectation values of $Q$
and $P$ and using (\ref{hsym}) as a classical Hamiltonian function
from which the canonical equations are derived. In this way one
obtains the coupled Bloch-oscillator equations representing the
dynamics of the system in the mixed description
\begin{eqnarray}\label{eommix}
\dot{x} &=& - \sqrt{2p}\,r\,Q\, y
\nonumber \\
\dot{y} &=&  \sqrt{2p}\,r\,Q\, x + z
\nonumber \\
\dot{z} &=& -y
\\
\dot{Q} &=& P
\nonumber \\
\dot{P} &=& -\,r^2 Q - \sqrt{p\over2}\,r z 
\nonumber
\end{eqnarray}
Besides the energy  
\begin{equation}
E = -{x\over2} + {1\over2}(P^2 + r^2Q^2) + \sqrt{p\over2}rQz  
\end{equation}
there is a second integral of the motion
restricting the flow associated with the quantum subsystem to the
surface of the unit radius Bloch sphere
\begin{equation}\label{bln}
R^2 = x^2 + y^2 + z^2 = 1\,.
\end{equation}
Sometimes it is advantageous to make use of this conserved quantity in
order to reduce the number of variables to four, e.\ g.\ when a
formulation in canonically conjugate variables is desired also for the
excitonic subsystem. One then introduces an angle $\phi$ by  
\begin{equation}\label{defphi}
x = \sqrt{1-z^2}\cos\phi\,,\hspace*{1cm}y=\sqrt{1-z^2}\sin\phi\,.
\end{equation}
We shall replace the usual Bloch variables by these coordinates 
where it is appropriate.

\subsection{ Adiabatic Basis}
In this case one first solves the eigenvalue problem of the part of
the Hamiltonian (\ref{hsym}) which contains excitonic operators
\begin{equation}\label{adham}
h^{{\rm (ad)}} = -{\sigma_x\over2} + \sqrt{p\over2}rQ\sigma_z \,,
\end{equation}
where $Q$ is considered as an adiabatic variable. The eigenvalues of
(\ref{adham}) are given by
\begin{equation}
\epsilon_\pm^{{\rm (ad)}}(Q)\, =  \pm {1\over2}w(Q),
\end{equation}
where
\begin{equation}\label{wu}
w(Q)\, = \sqrt{1 + 2pr^2 Q^2}.
\end{equation}
The eigenvalues are part of the adiabatic potentials for the slow 
subsystem
\begin{equation}\label{adpot}
U_\pm^{{\rm (ad)}}(Q) = {1\over 2} r^2Q^2 + \epsilon_\pm^{{\rm (ad)}}(Q)  \, ,
\end{equation}
The two eigenstates $(\alpha=1,2)$ of (\ref{adham}) represented by 
the fixed basis are given by
\begin{equation}\label{adbv+}
|\alpha=2,Q\rangle = {1\over\sqrt{2}}(-\sqrt{1 + c(Q)}\,|2\rangle+\sqrt{1 - c(Q)}\,|1\rangle)
\end{equation}
and 
\begin{equation}\label{adbv-}
| \alpha=1,Q \rangle = {1\over\sqrt{2}}(+\sqrt{1 - c(Q)}\,|2\rangle+\sqrt{1 + c(Q)}\,|1\rangle)
\end{equation}
with
\begin{equation}
c(Q) := {\sqrt{2p}rQ\over w(Q)}\,.
\end{equation}
The state vector of the excitonic subsystem is expanded in the
adiabatic basis $|\psi\rangle = \sum_{\alpha}c^{{\rm (ad)}}_\alpha |\alpha,Q \rangle $ and
inserted into the time dependent Schr\"odinger equation. For obtaining
the complete evolution equations in the adiabatic basis one has to
take into account the time derivative of the expansion coefficients
$c^{{\rm (ad)}}_\alpha$ as well as the nonadiabatic couplings due to the time
dependence of the states $|\alpha,Q(t)\rangle$.  The neglect of these
couplings would result in the adiabatic approximation. Using $(d/d t)
|\alpha,Q \rangle = \dot Q (d/d Q)|\alpha,Q\rangle$ the nonadiabatic coupling
function
\begin{equation}\label{ncg}
\varphi_{\alpha\beta} := \left\langle\alpha,Q\left|{\partial\over\partial Q}\right|\beta,Q \right\rangle,
\end{equation}
$(\varphi_{\alpha\beta}=-\varphi_{\beta\alpha})$ is found, which in
case of the eigenstates (\ref{adbv+},\ref{adbv-}) is explicitly given
by
\begin{equation}\label{nce}
\varphi_{12} = - \frac{\sqrt{p}r}{\sqrt{2}[w(Q)]^2}\,.
\end{equation}
Introducing now analogously to (\ref{bl}) the Bloch variables in the
adiabatic basis and treating the oscillator in the classical
approximation one obtains the coupled Bloch-oscillator equations in
the adiabatic basis
\begin{eqnarray}\label{eomad}
\dot x^{{\rm (ad)}} &=& 2P\, \varphi_{12}(Q)\, z^{{\rm (ad)}} - w(Q)\, y^{{\rm (ad)}}\nonumber \\ 
\dot y^{{\rm (ad)}} &=& w(Q)\, x^{{\rm (ad)}}\nonumber \\ 
\dot z^{{\rm (ad)}} &=& -2P\, \varphi_{12}(Q)\, x^{{\rm (ad)}}\\
\dot Q &=& P\nonumber \\ 
\dot P &=& -r^2Q + \sqrt{p}\,w(Q)\varphi_{12}\,x^{{\rm (ad)}} - \sqrt{p\over2}\,r\,c(Q)\,z^{{\rm (ad)}}\nonumber
\end{eqnarray}
The energy is now expressed by the adiabatic Bloch variable $z^{{\rm (ad)}}$
\begin{equation}\label{ad1dz}
E = w(Q){z^{{\rm (ad)}}\over2} + {1\over 2}\left(P^2 + r^2Q^2\right)
\end{equation}
and the flow is again located on the surface of the unit Bloch sphere.
Neglecting the nonadiabatic couplings $\varphi_{12}=0$ one obtains the
dynamics of the decoupled adiabatic oscillators. The adiabatic
oscillators can be considered as one dimensional integrable subsystems
corresponding to the Hamiltonians
\begin{equation}\label{ad1d}
h_\pm^{{\rm (ad)}} = {1\over 2} P^2 + U_\pm^{{\rm (ad)}}(Q) \, , 
\end{equation}
where $U_\pm^{{\rm (ad)}}(Q)$ is given by (\ref{adpot}). 
The connection between the Bloch variables in the fixed and the adiabatic
basis is given by\\[3mm]
\begin{tabular}{ll}
$ x = - c(Q) x^{{\rm (ad)}} - \sqrt{1-c(Q)^2}\,z^{{\rm (ad)}}  \hspace*{1cm}$&
$ x^{{\rm (ad)}} = - c(Q) x - \sqrt{1-c(Q)^2}\,z $\\
$ y = -y^{{\rm (ad)}} $&
$ y^{{\rm (ad)}} = -y $\\
$ z = c(Q) z^{{\rm (ad)}} - \sqrt{1 - c(Q)^2} x^{{\rm (ad)}} $&  
$ z^{{\rm (ad)}} = c(Q) z - \sqrt{1 - c(Q)^2} x $  
\end{tabular}
\begin{equation}\label{trbloch}
\sqrt{1-z^2}\,\sin\phi=-\sqrt{1-(z^{{\rm (ad)}})^2}\,\sin\phi^{{\rm (ad)}}
\end{equation}
Using these transformation formulas one can show that the equations of
motion (\ref{eommix}) derived in the fixed basis are actually
equivalent to those in the adiabatic basis (\ref{eomad}).
 
\subsection{Fixed points and bifurcation}
Essential information about the phase space of the excitonic-vibronic
coupled dimer is contained in the location and the stability
properties of the fixed points of the mixed quantum-classical
dynamics.  Setting in the equations of motion for the fixed basis
(\ref{eommix}) all time derivatives to zero, we find for any
stationary state
$$
Q_s = -{1\over r}\sqrt{p\over2}z_s \,,\hspace*{1cm} P_s = 0 \,,\hspace*{1cm} y_s = 0
$$
\begin{equation}\label{statcon}
z_s - p x_s z_s = 0 \,. 
\end{equation}
The stability properties of a fixed point are determined by a
linearization of the equations of motion using canonical variables
\cite{ES}.

It is appropriate to subdivide all stationary points according to
whether they are located in the bonding region $x_s > 0$ or in the
antibonding region $x_s < 0$. There is no transition between these two
groups when the parameters of the system are varied since $x_s = 0$ is
excluded by (\ref{statcon}). This terminology is in accordance with 
molecular physics where it is common to refer to the state $x=1$ with
symmetric site occupation amplitudes $c_1 = c_2$ as bonding and to the
state $x=-1$ with antisymmetric amplitudes $c_1=-c_2$ as antibonding.

\subsubsection{Bonding region ($x_s > 0$)}
We consider the bonding region first. The location of the fixed points
is obtained from (\ref{statcon}) using the additional restriction
\begin{equation}
x_s^2 + z_s^2 = 1\,.
\end{equation}
One finds the following solutions in dependence on the value of the
dimensionless coupling strength $p$:
\begin{description}
\item{(A) $0 \le p \le 1$:}
  In this case (\ref{statcon}) allows for a single solution only.
  \begin{equation} {\bf g:}\hspace*{1cm} 
  x_s = 1 \,,\hspace*{1cm} z_s = 0 \,,\hspace*{1cm} Q_s = 0 \,,\hspace*{1cm} E_s =-{1\over 2} 
  \end{equation}
  This point is the bonding ground state corresponding to a symmetric
  combination of the excitonic amplitudes $c_1 = c_2 = 1/\sqrt{2}$.
  $\bf g$ is stable elliptic. 
\item{(B) $p \ge 1$:}
A bifurcation has occurred and we obtain three stationary points.
\begin{equation}
\begin{array}{ccc}
{\bf g_{\pm}:} & x_s = {1\over p} & z_s = \pm{\sqrt{p^2-1}\over p} \\
& Q_s=\pm{\sqrt{p^2-1}\over\sqrt{2p\,}r} & E_s= -{p^2+1\over4p} 
\end{array}
\end{equation}
These two points are stable elliptic.  
\begin{equation} 
{\bf h:}\hspace*{1cm}
x_s = 1\,,\hspace*{1cm} z_s =0 \,,\hspace*{1cm} Q_s = 0 \,,\hspace*{1cm} E_s = -{1\over 2} 
\end{equation}
The point $\bf h$ is at the position of the former ground state, but in
contrast to $\bf g$ it is unstable hyperbolic. 
\end{description}
 
\begin{figure}
\centerline{
\psfig{figure=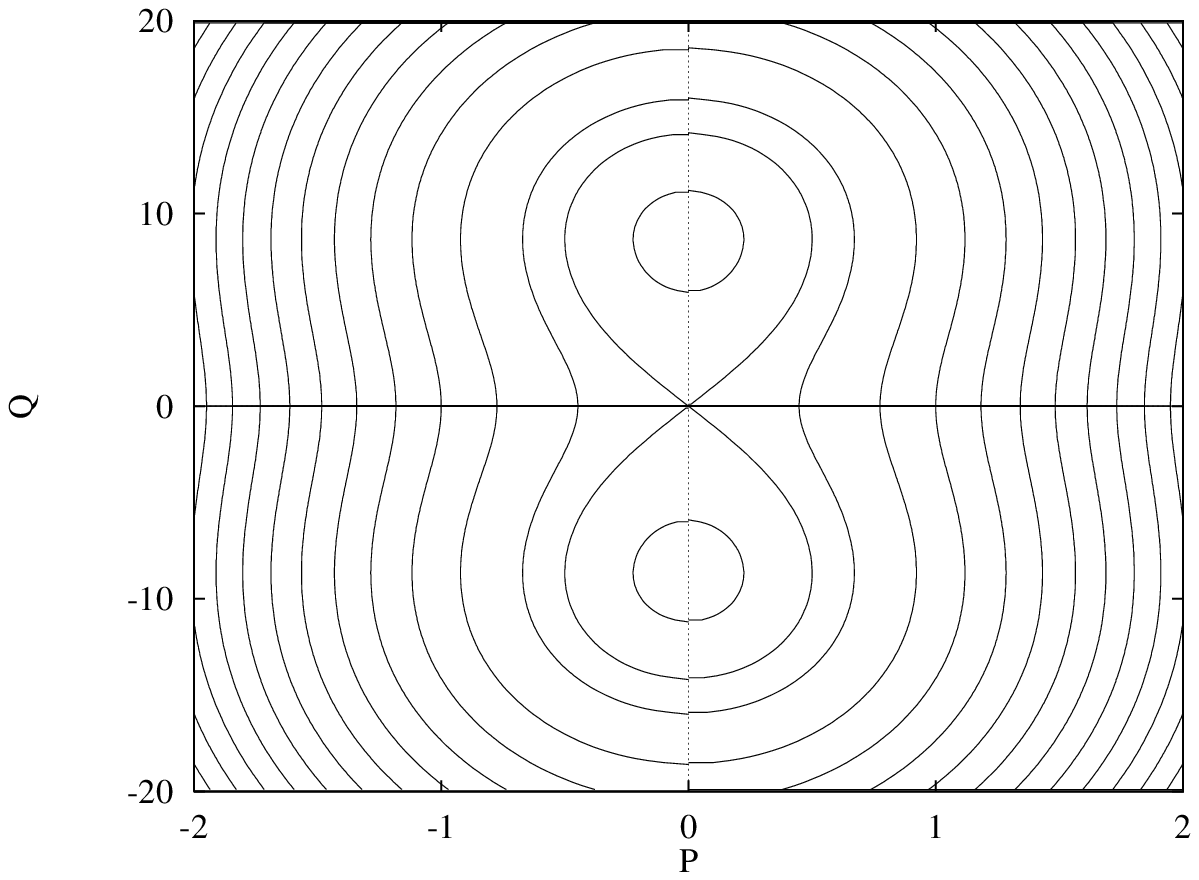,height=5cm,angle=0}
\psfig{figure=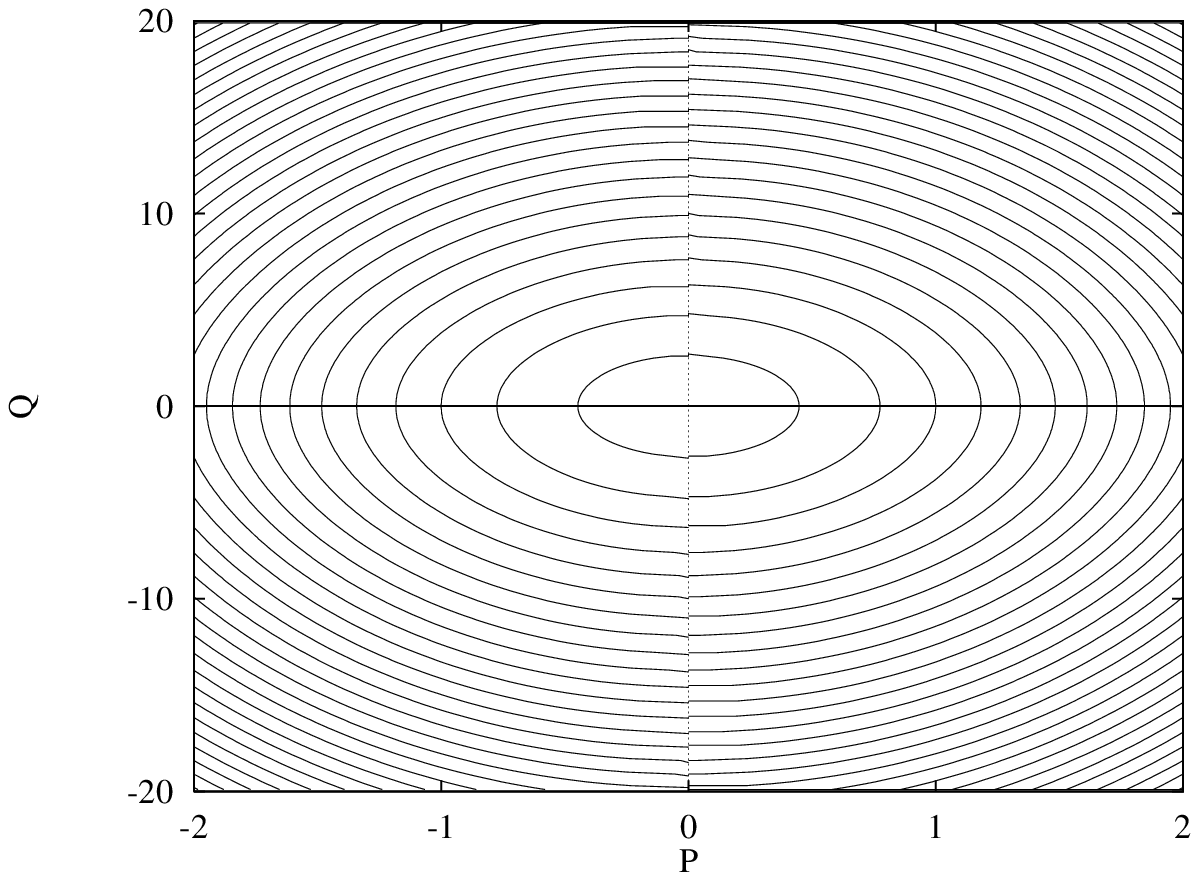,height=5cm,angle=0}
} 
\caption{\label{adpotplot}
  Phase space plots for the adiabatic oscillators (left: lower
  potential, right: upper potential) for $p=2$ and $r=0.1$.  
}
\end{figure}

\noindent
The parameter $p$ governs a pitchfork bifurcation: The ground state g
below the bifurcation ($p < 1$) splits into two degenerate ground
states $\bf g_\pm$ above bifurcation ($p > 1$). At the former ground state
a hyperbolic point $\bf h$ appears. This situation is also obvious from
fig.~\ref{adpotplot}(a).

\subsubsection{Antibonding region ($x_s < 0$)}
Independent on the coupling strength $p$ we have in this region only one 
solution of (\ref{statcon}) (see fig.\ \ref{adpotplot}(b)):
\begin{equation} 
{\bf e:} 
z_s = 0 \,,\hspace*{1cm} x_s= -1 \,,\hspace*{1cm} Q_s = 0 \,,\hspace*{1cm} E_s = +{1\over 2} 
\end{equation}
This stationary state corresponds to an
antisymmetric combination of the excitonic amplitudes $c_1 = -c_2 =
1/\sqrt{2}$. $\bf e$ is stable for
\begin{equation}\label{stababond}
\frac{|r^2 - 1|}{r} > 2\sqrt{p}\,,
\end{equation} 
which holds when the system is not in resonance and in particular for
the adiabatic case $r \ll 1$.

\paragraph*{}
Since the equations of motion in the fixed and in the adiabatic basis
are equivalent, it is clear that the fixed same points (\ref{statcon})
can also be obtained from (\ref{eomad}).  Setting in (\ref{eomad}) the
time derivatives of $x$, $y$ and $Q$ equal to zero, one finds for the
stationary states of the adiabatic case
\begin{eqnarray}
\label{ans}
x_s^{{\rm (ad)}} = 0 \,,\hspace*{1cm}
y_s^{{\rm (ad)}} = 0 \,,\hspace*{1cm}
P_s^{{\rm (ad)}} = 0\,,
\end{eqnarray}
leaving for the stationary values of $z_s^{{\rm (ad)}}$ the poles  
\begin{equation}\label{adz}
z_s^{{\rm (ad)}} = \pm 1
\end{equation}
It is worth noting that in the adiabatic basis the stationary states
are always located at $z^{{\rm (ad)}}_s=\pm 1$ and this will be the case for any
system treated in mixed quantum-classical description and restricted
to the two lowest adiabatic levels. A specific feature of using the
adiabatic basis is the independence of the location of the fixed
points on the explicit form of the nonadiabatic coupling function
since this function enters the equations of motion in form of the
products $\varphi_{12}(Q)P$ and $\varphi_{12}(Q)x^{{\rm (ad)}}$ which 
drop out at a fixed point because of (\ref{ans}) and (\ref{adz}).
From (\ref{trbloch}) it is moreover easy to see that the fixed points
in the bonding region are located within the lower adiabatic potential
while the antibonding fixed points belong to the upper one.
The eq.\ $\dot {P}=0$ reduces to
\begin{equation}\label{adQ}
(w(Q) + pz_s^{{\rm (ad)}})Q=0\,.
\end{equation}    
For $z^{{\rm (ad)}}_s=+ 1$ the only solution of (\ref{adQ}) is $Q_s=0$, whereas
for $z^{{\rm (ad)}}_s=-1$ one obtains additional solutions for $p>1$. These
solutions are easily seen to correspond to the bifurcation discussed
above.

\subsection{Integrable Approximations}
Before we investigate the dynamics of the complete coupled equations
of motion (\ref{eommix}) or (\ref{eomad}) we would like to mention two
integrable approximations to the model. The first and trivial
integrable approximation is to set in the equations of motion in the
fixed basis (\ref{eommix}) $p=0$ which results in a decoupling of the
excitonic and vibronic motions. The second and more interesting
integrable approximation is obtained by neglecting the nonadiabatic
coupling function $\varphi_{12}=0$ in the equations of motion
(\ref{eomad}) of the adiabatic basis which defines the adiabatic
approximation from a dynamical point of view.  In this approximation
some of the nonlinear features of the model are still contained in the
integrable adiabatic reference oscillators (\ref{ad1d}). In particular
the lower adiabatic potential (\ref{adpot}) displays the bifurcation
from a single minimum structure to the characteristic double well
structure when the parameter $p$ (\ref{par}) passes through the
bifurcation value $p=1$.  It is also important to note that the fixed
point structure is not changed when the nonadiabatic couplings are
switched off: Neglecting $\varphi_{12}$ in (\ref{eomad}) results in
the same fixed points equations as in the case including the
nonadiabatic couplings.  From the formal side the neglect of
$\varphi_{12}$ not necessarily leads to $z_s^{{\rm (ad)}} = \pm 1$: According to
the equations of motion(\ref{eomad}) $\varphi_{12}=0$ implies $z_s^{{\rm (ad)}}
= const$.  Then in the dynamics of the adiabatic approximation both
adiabatic modes can be occupied and only the transitions between them
are switched off. The oscillator equations become autonomous
describing regular motions according to the classical Hamilton
function (\ref{ad1dz}) with $z_s^{{\rm (ad)}}$ as a parameter.  The oscillator
coordinate $Q(t)$ enters the Bloch equations for $x(t)$ and $y(t)$.
The equations for the latter describe the regular motion on a circle
generated by an intersection of the Bloch sphere with the plane
$z_s^{{\rm (ad)}} = const$ on which the phase oscillations between the modes are
realized.

In the following we demonstrate that the regular structures associated
with the adiabatic approximation are present in both the mixed and
fully quantized descriptions. At the same time we show that the
complete coupled system of Bloch-oscillator equations, i.~e.\ 
including the nonadiabatic couplings, displays dynamical chaos. This
identifies the nonadiabatic couplings as a source of nonintegrability
and chaos in the mixed description of the system and rises the
question about the signatures of this chaos after full quantization
is performed. The latter problem will be addressed in the last
section.

\subsection{Dynamical Properties}
\label{sec:dynmix}
The dynamical properties of the coupled Bloch-oscillator equations
(\ref{eommix}) were analyzed by a direct numerical integration. Some of
our results, such as the presence of chaos in the mixed description of
the excitonic-vibronic coupled dimer, were reported in \cite{ES}.
Therefore the aim of this section is twofold: On the one side we
reconsider the findings in \cite{ES} relating the dynamical structures
to the adiabatic approximation, in which the integrable reference
systems (\ref{ad1d}) can be defined. This clarifies the role of the
nonadiabatic couplings in the formation of the dynamics of the model,
which was not done before. On the other side we provide the necessary
characterization of the phase space structure, such as the location of
the separatrix dividing the phase space into trapped and detrapped
solutions, and the identification of the regions and associated
parameters belonging to the regular and chaotic parts of the dynamics,
respectively.  The latter points will provide the basis to perform the
comparison of the mixed description with the full quantum evolution in
the next section.

The numerical integration of the equations of motion in the mixed
description can be performed both in the fixed (\ref{eommix}) and the
adiabatic basis (\ref{eomad}).  The integration in the fixed basis,
however, is numerically simpler and the fixed Bloch variables provide
a more convenient frame for the excitonic motion. Hence we used the
fixed basis for a numerical integration. It must be stressed, however,
that the representation of the excitonic system in the fixed and the
adiabatic basis are equivalent, if the nonadiabatic couplings are
included.  The connection between both representations is given by the
eqs. (\ref{trbloch}).

In view of the existence of two integrals of the motion three
variables from the total of five variables of the system are
independent.  Therefore a standard two dimensional Poincar\'e surface
of section is defined by fixing one variable.  According to the choice
of variables sections can be defined for both the oscillatory and
excitonic subsystems.  The dynamics of the system was found to depend
crucially on the choice of the total energy with respect to the
characteristic energies of the system such as the minima of the
adiabatic potentials and above the bifurcation the energy
corresponding to the hyperbolic fixed point $E_h$.  Above the
bifurcation the separatrix structure, which divides the phase space into
characteristic parts, is present. The regular phase space structure
following from the integrable adiabatic reference oscillators
(\ref{ad1d}) above the bifurcation is shown in fig.\ \ref{adpotplot}.

\begin{figure}
\centerline{
\vspace*{3mm}
\psfig{figure=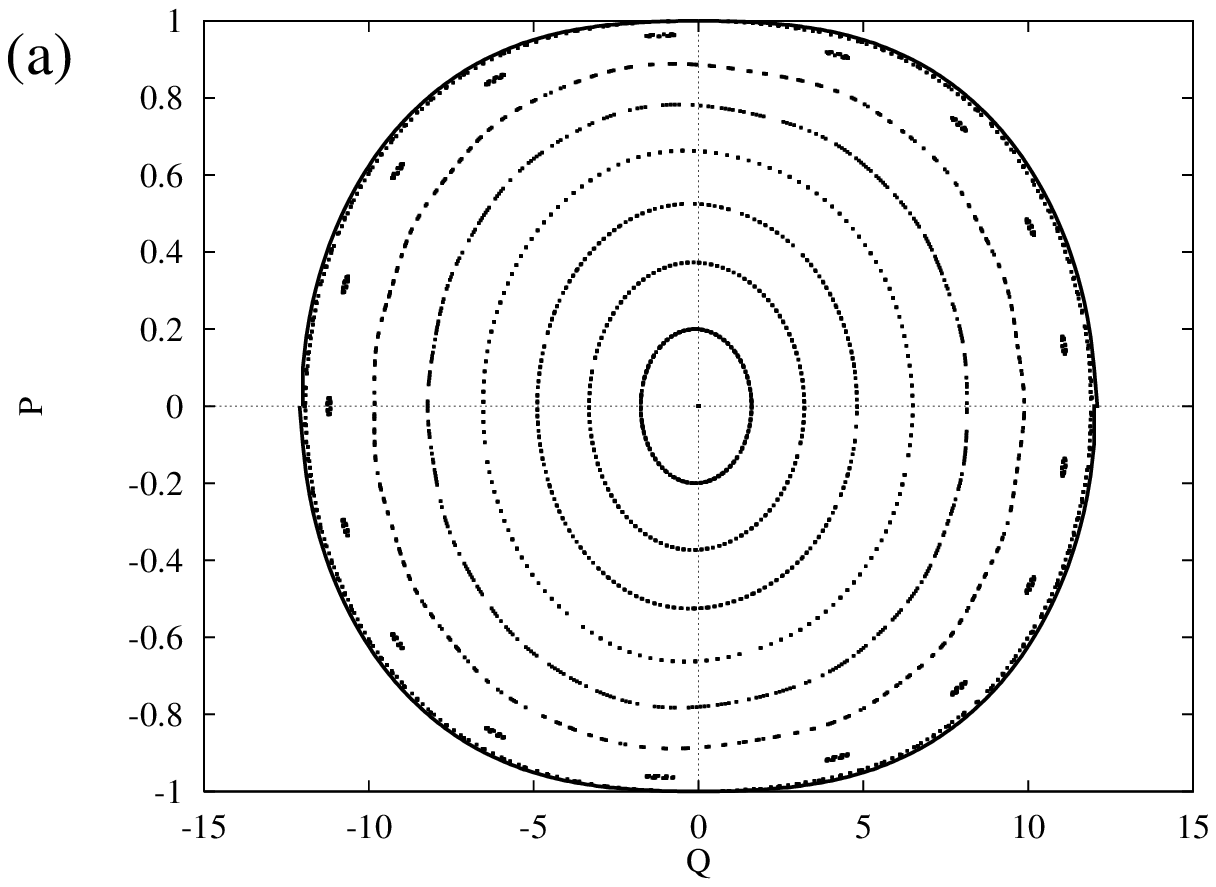,height=5cm,angle=0}
\psfig{figure=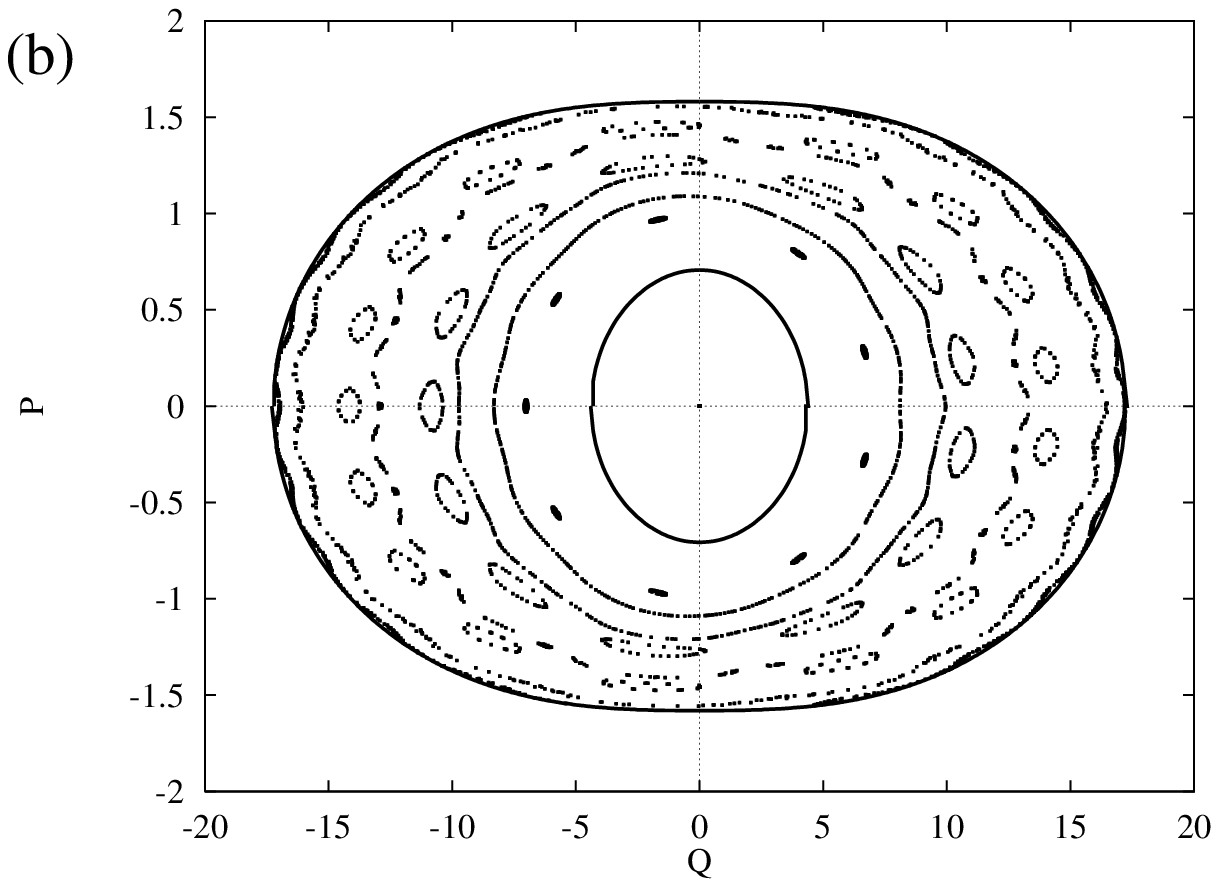,height=5cm,angle=0}
} 
\caption{\label{pq-0.8}
  Poincar\'e section in the oscillator variables for $p = 0.8$ (below
  bifurcation), $r = 0.1$ and (a) $E = 0.0$ and (b) $E = 0.75$. The
  Bloch variable $y$ is fixed ($y = 0$, $dy/dt > 0$).  }
\end{figure}

In fig.\ \ref{pq-0.8}(a) a Poincar\'e section in oscillator variables
is presented for the value $p=0.8$ which is below the bifurcation
value $p=1$. In this case the adiabatic potential $U_-(Q)$ has a
single minimum. The total energy is chosen at $E=0$, i.\~e.\ well
below the minimum of the upper adiabatic potential. Therefore the
influence of this potential is small and the oscillator dynamics can
be expected to be close to the regular dynamics of the lower reference
oscillator associated with $U_-(Q)$. This is indeed confirmed by fig.\ 
\ref{pq-0.8}(a). There is, however, a chain of small resonance islands
in the outer part of the section, which is due to resonance between
the oscillator motion and the occupation oscillations between the
adiabatic modes corresponding to the finite $\dot z^{{\rm (ad)}}$ in the case of
the presence of the nonadiabatic couplings.  The interaction between
the occupation oscillations due to the finite $\dot z^{{\rm (ad)}} $ and the
oscillator motion becomes much more pronounced for higher energies.  A
corresponding Poincar\'e section is displayed in fig.\ \ref{pq-0.8}(b)
for the same value $p=0.8$ of the coupling constant, but with the
energy now chosen above the minimum of the upper adiabatic potential.
This choice of the energy allows according to (\ref{ad1dz}) for a much
broader range for the variation of the variable $z^{{\rm (ad)}}$ and
consequently the nonadiabatic couplings are more effective.
Correspondingly we observe now several resonance chains.

\begin{figure}
\centerline{
\vspace*{3mm}
\psfig{figure=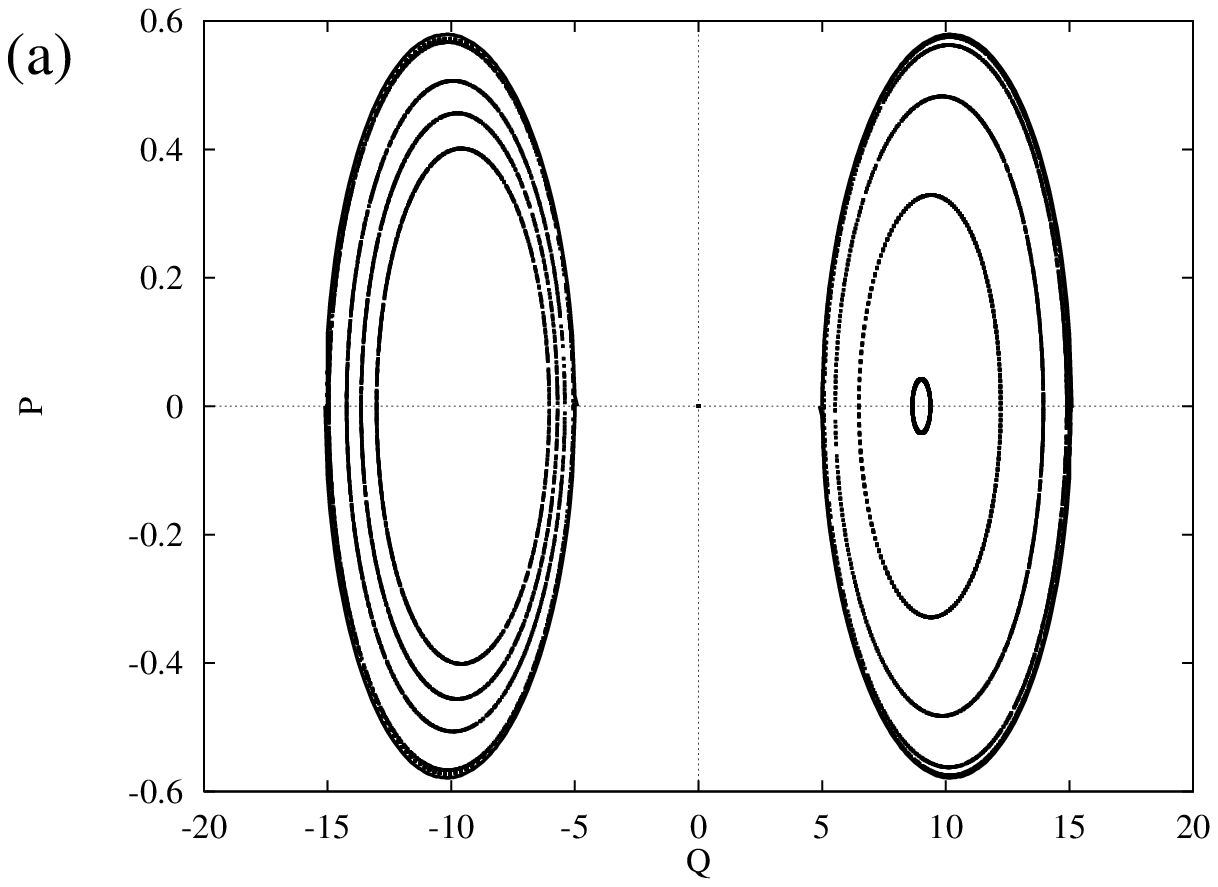,height=5cm,angle=0}
\psfig{figure=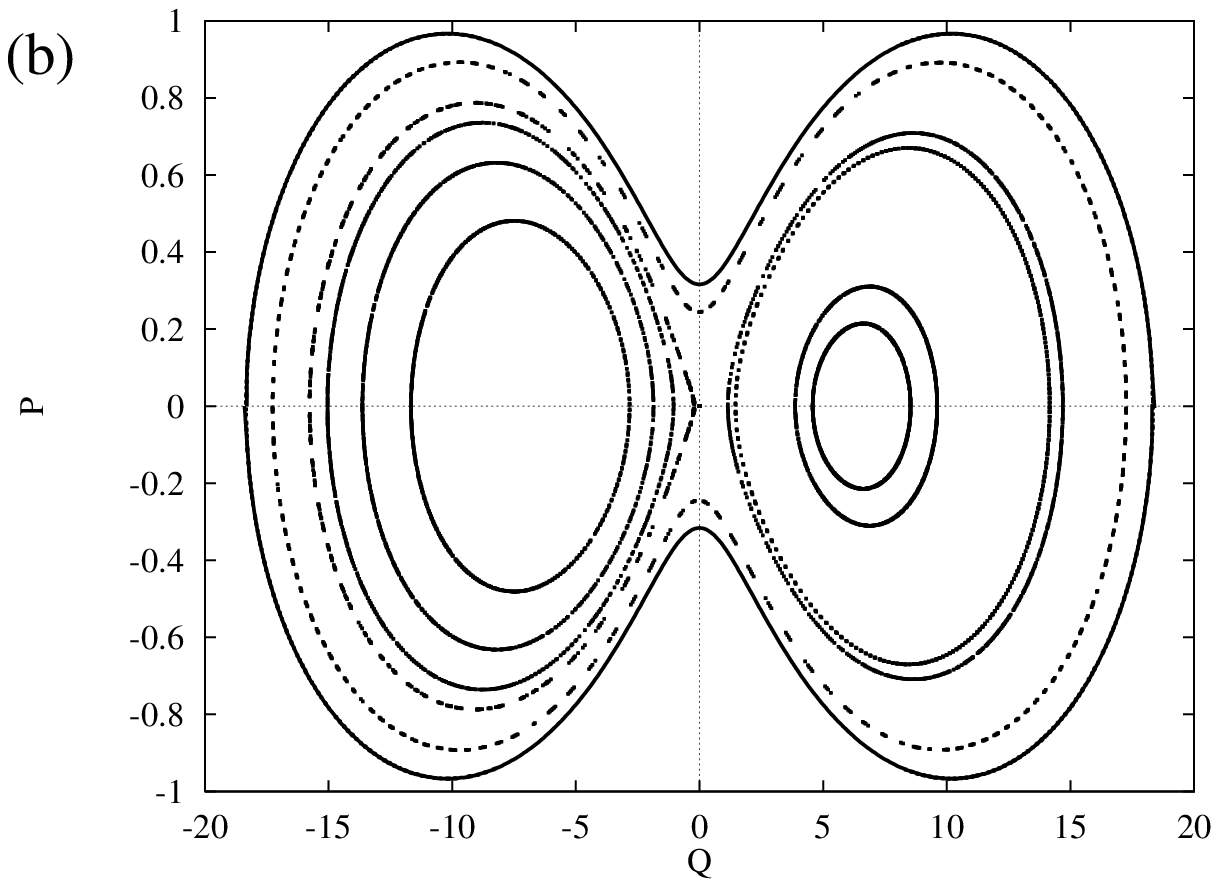,height=5cm,angle=0}
}
\centerline{
\vspace*{3mm}
\psfig{figure=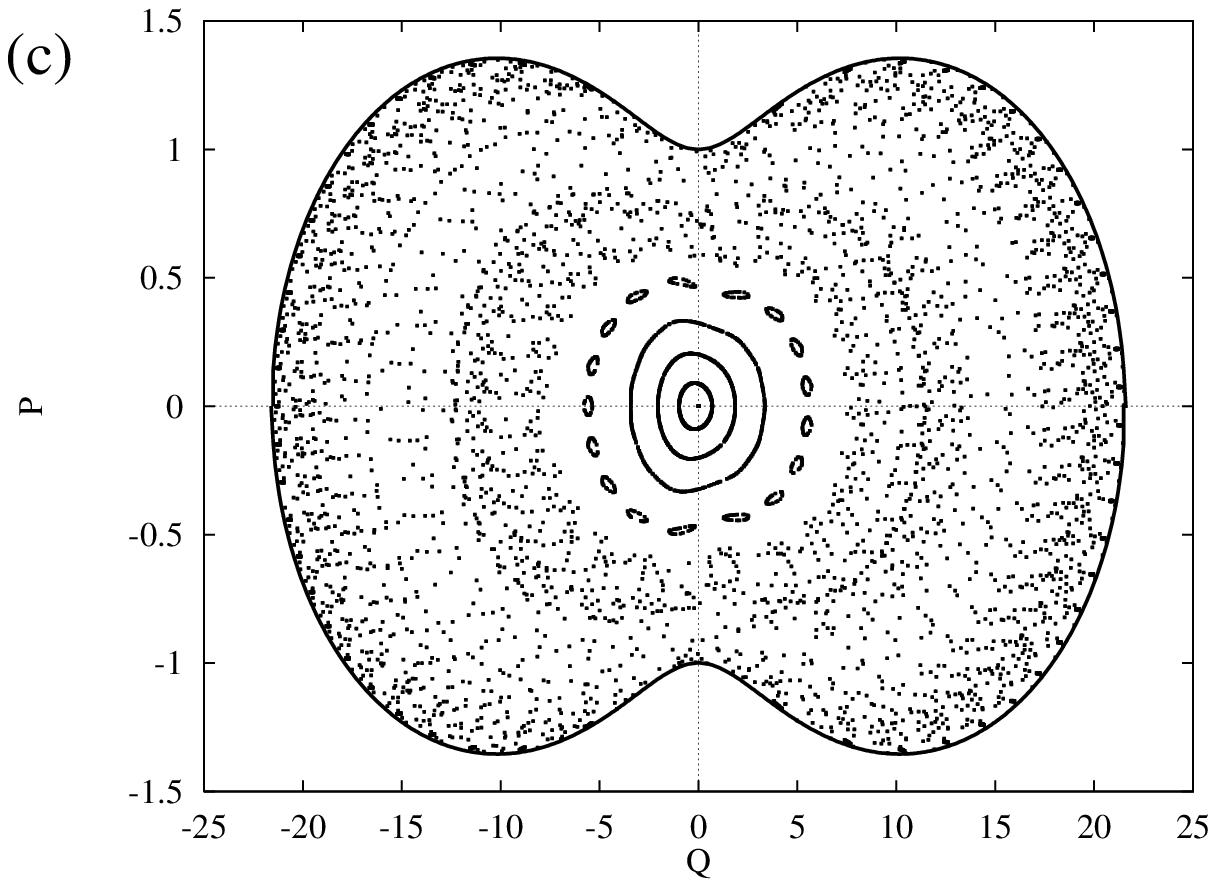,height=5cm,angle=0}
\psfig{figure=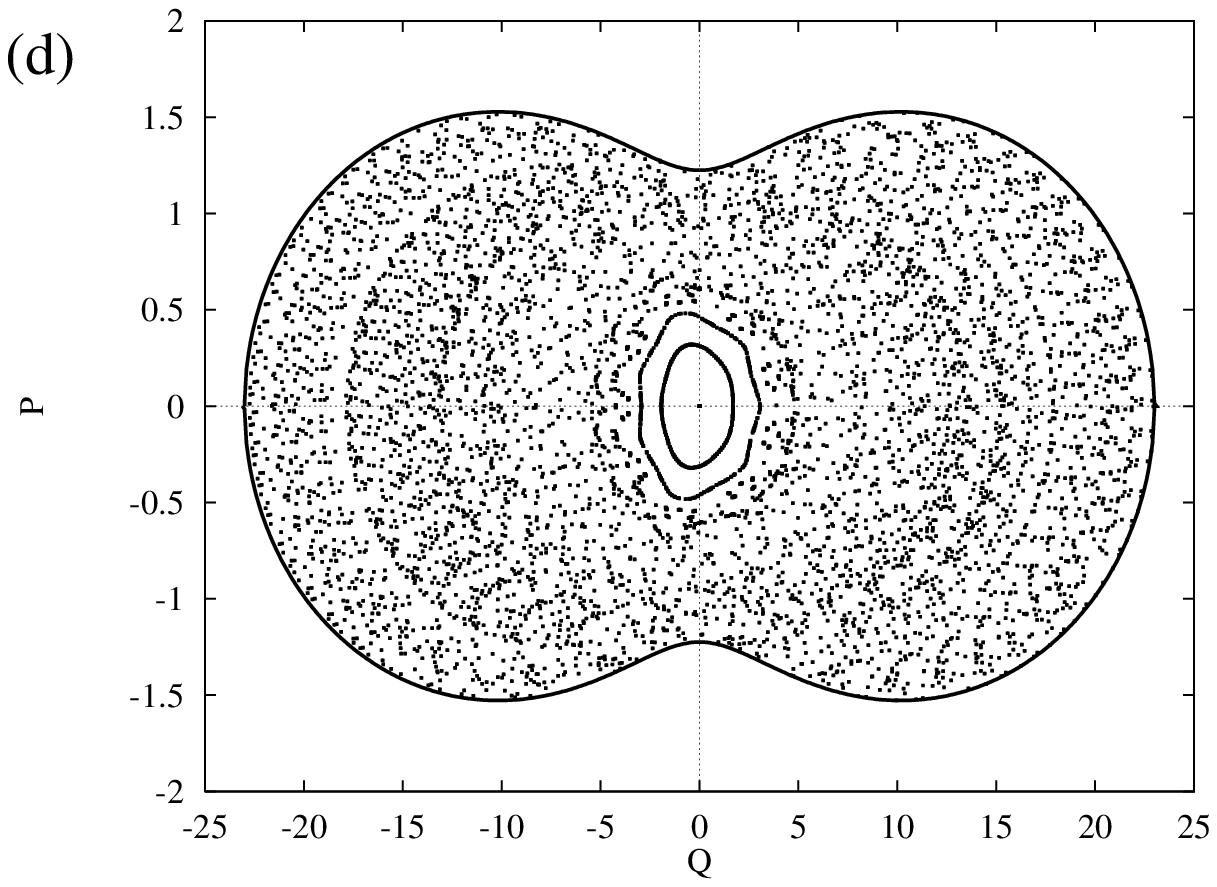,height=5cm,angle=0}
} 
\caption{\label{pq-3.4}
  Poincar\'e sections in the oscillator variables for $p = 3.4$ (above
  bifurcation), $r = 0.1$ and (a) $E = -0.75$, (b) $E = -0.45$, (c) $E
  = 0.0$ and (d) $E = 0.25$. The Bloch variable $y$ is fixed ($y = 0$,
  $dy/dt > 0$).}
\end{figure}

Increasing the coupling above the bifurcation value $p=1$, but fixing
the total energy below $E_h$ one expects regular oscillations around
the displaced minima of the double well structure in $U_-(Q)$. A
Poincar\'e section in the oscillator variables corresponding to this
behavior is shown in fig.\ \ref{pq-3.4}(a), where $p=3.4$ and the
total energy is below the saddle point of the potential $U_-(Q)$.
Increasing the energy to a value slightly above $E_h$ one finds
sections displaying oscillations resembling the separatrix structure
as shown in fig.\ \ref{pq-3.4}(b).

For energies well above $E_h$ chaotic trajectories do exist.
Characteristic examples are provided by the Poincar\'e sections in
oscillator variables displayed in fig.\ \ref{pq-3.4}(c) and (d), where
the regions of regular and chaotic behavior of the oscillator
subsystem are shown for two cases of total energy above $E_h$. In the
case (c) the total energy is below the case (d). It is seen how with
increasing energy the regular part of the oscillator phase space
becomes smaller and the chaotic part increases.  The corresponding
regular and chaotic components of the excitonic subsystem are located
in the antibonding and bonding regions of the Bloch sphere,
respectively (see also fig.\ \ref{03} below).  Relating the location
of the dynamics on the Bloch sphere to the energy of the excitonic
subsystem we find that for chaotic trajectories the excitonic
subsystem is in an energetically low state within the bonding region
of the Bloch sphere whereas for regular trajectories the excitonic
subsystem is in its energetically high state within the antibonding
region. Correspondingly, the energy of the vibronic subsystem is high
for chaotic dynamics and low for regular dynamics, because the total
energy is the same for all the trajectories displayed in each of the
figures \ref{pq-3.4}(c) and (d).  Hence the destruction of the regular
dynamics is connected with the energy of the vibronic subsystem:
Regular dynamics is realized for oscillator states with low energy,
small amplitude oscillations and consequently small effective
coupling, whereas high oscillator energy destroys the regular
structures and results in global chaos.

The dynamics of the oscillator subsystem is complemented by the
Poincar\'e sections on the surface of the Bloch sphere showing the
behavior of the excitonic subsystem. In the figs.\ \ref{03}(a)-(d)
such a typical set of Poincar\'e sections is presented for different
energies and above the bifurcation ($p = 2.0$). The sections
correspond to the left turning point of the oscillator. For low energy
one finds regular dynamics in the region of the bifurcated ground
states.  These regular trajectories represent the self trapped
solutions of the system in which the exciton is preferentially at one
of the sites of the dimer and correspond to the one sided oscillations
of the vibronic subsystem of fig.\ \ref{pq-3.4}(a).  Increasing the
energy local chaos starts in the vicinity of the hyperbolic point
$E_h$.  The local chaos can be considered as a perturbation of the
dynamics near the saddle of the lower potential $U_-(Q)$ due
the nonadiabatic couplings of the adiabatic oscillators.  With
increasing energy chaos spreads over the Bloch sphere leaving only
regular islands in the region of antibonding states associated with
the upper adiabatic potential and in accordance with the dynamics of
the vibronic subsystem discussed above. For high enough energy the
coupling between the adiabatic reference oscillators almost completely
destroys regular structures and results in global chaos.

\section{Quantum Evolution}\label{qevol}
We now turn to the dynamics in the full quantum description of the
model considering in the Hamiltonian (\ref{hsym}) the coordinate $Q$
and the momentum $P$ as non-commuting quantum variables.  We focus on
the features of the evolution in the adiabatic parameter region for
$r\ll 1$.  The evolution of the full quantum state vector of the
system satisfying some fixed initial condition is computed from the
eigenstate representation of the Hamiltonian.  For a realistic
description of the system in the adiabatic parameter region a large
number of eigenstates had to be used in the expansion.
Correspondingly, the diagonalization of the Hamiltonian (\ref{hsym})
with $Q$ and $P$ being quantum operators was performed using a large
set of oscillator eigenfunctions for the undisplaced oscillator as a
basis, i.~e.\ the basis was constructed from the product states
$|n,\nu\rangle:=|n\rangle \otimes |\nu\rangle$, where the index $n=1,2$ labels the
two sites of the dimer and $\nu=0,1,\dots$ stands for the oscillator
quantum number.  In this basis the quantized version of the
Hamiltonian (\ref{hsym}) is represented by the matrix
\begin{equation}
\langle n,\nu| H | n',\nu '\rangle =
-{[1-(-1)^{\nu+\nu'}]\over 4}\delta_{\nu ,\nu'} \,+\, 
r \left( \nu + {1 \over 2}\right)\delta_{n, n'}\delta_{\nu,\nu '} \,+
\end{equation}
\begin{equation}\label{matel}
+\, 
{\sqrt{p\,r}\over 2}(-1)^n\,(\sqrt{\nu'}\delta_{\nu,\nu ' -1} + 
\sqrt{\nu}\delta_{\nu ,\nu ' + 1})\,\delta_{n, n'}\,. 
\end{equation}
The typical number of oscillator eigenfunctions used was $750$
yielding a total of $1500$ basis states. 
The properties of the stationary eigenstates, the fine structure of
the spectrum and in particular the influence of the adiabatic
reference oscillators and the role of the nonadiabatic couplings in
the formation of the spectrum were reported in \cite{inc}. Here we
consider the nonstationary properties of the full quantum system based
on this eigenstate expansion and demonstrate how the nonlinear
features of the dynamics in the mixed quantum-classical description
are reflected in the time dependence of the full quantum state vector
$|\Psi(t)\rangle$.

We investigated the evolution of wave packets initially prepared in
the product state
\begin{equation}\label{wp}
|\Phi,\alpha\rangle = |\Phi_{z_0,\phi_0}\rangle \otimes |\alpha_{Q_0,P_0}\rangle
\end{equation}  
where $\Phi$ is an excitonic two component wave function which is
specified up to an irrelevant global phase by the expectation values
of the Bloch variables $z$ and $\phi$ (see (\ref{defphi})).  $\alpha$
represents a standard coherent state in the oscillator variables,
which is specified by the complex parameter
\begin{equation}\label{aqp}
\alpha(Q,P) = \sqrt{r\over2}\langle\alpha|\hat Q|\alpha\rangle + 
{i\over\sqrt{2r}}\langle\alpha|\hat P|\alpha\rangle
\end{equation}
with $Q$ and $P$ being the corresponding expectation values of
position and momentum.

In order to map the motion of the full state vector $|\Psi(t)\rangle$ 
constructed from the eigenstate expansion according to the initial 
condition onto an analogue of the phase space of the mixed 
description, in which the oscillator is treated classically,  
we used for the oscillator subsystem the Husimi distribution, 
which is an appropriate quantum analogue to the classical 
phase space distribution (see e.\ g.\ \cite{Tak}). It is defined 
by projecting $|\Psi(t)\rangle$ on the manifold of coherent states  
\begin{equation}\label{husdist}
h_{z,\phi}(Q,P):=|\langle\Phi_{z,\phi},\alpha_{Q,P}|\Psi(t)\rangle|^2, 
\end{equation} 
where now $Q$ and $P$ are varied in the oscillator plane while 
$z$ and $\phi$ are fixed parameters.

Without the interaction between the subsystems a wave packet prepared
in a coherent oscillator state would travel undistorted along the
classical trajectory started at $(Q_0,P_0)$. A weak coupling below the
bifurcation ($p<1$) results in a similar picture (not displayed) with
the wave packet after some initial period almost uniformly covering
the classical trajectories such as those displayed in figs.\ 
\ref{pq-0.8}(a) and (b).
  
\begin{figure}
\centerline{
\psfig{figure=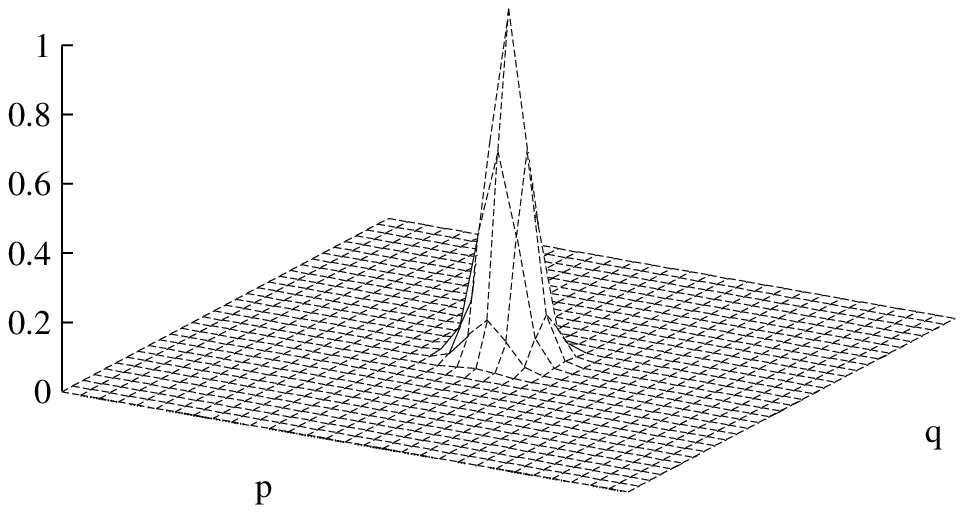,width=75mm}
\psfig{figure=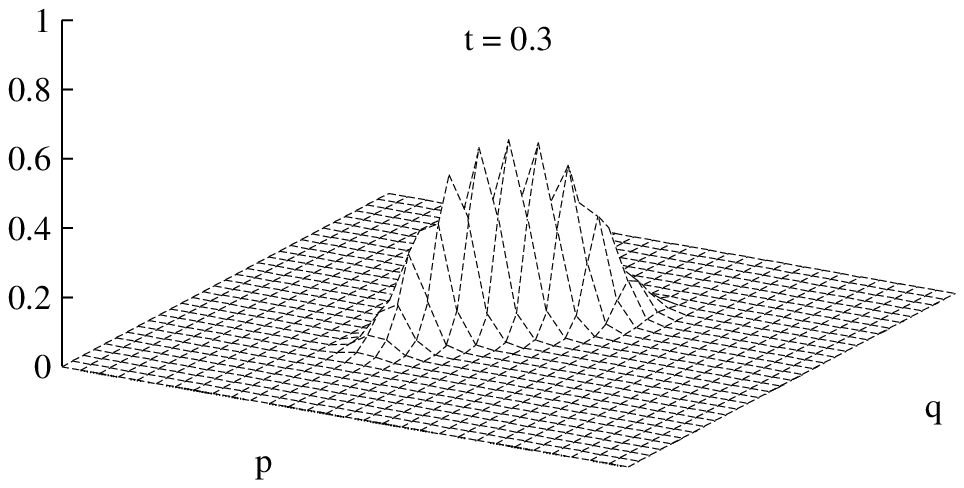,width=75mm}
}
\centerline{
\psfig{figure=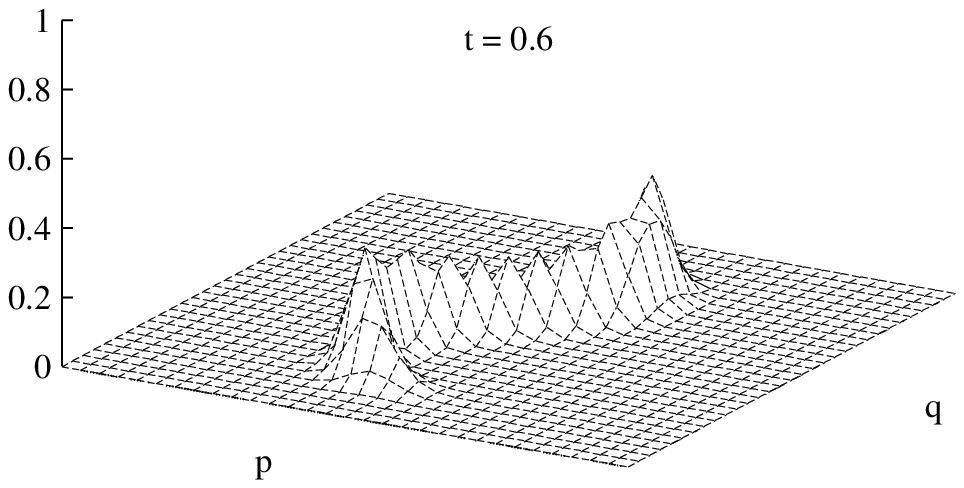,width=75mm}
\psfig{figure=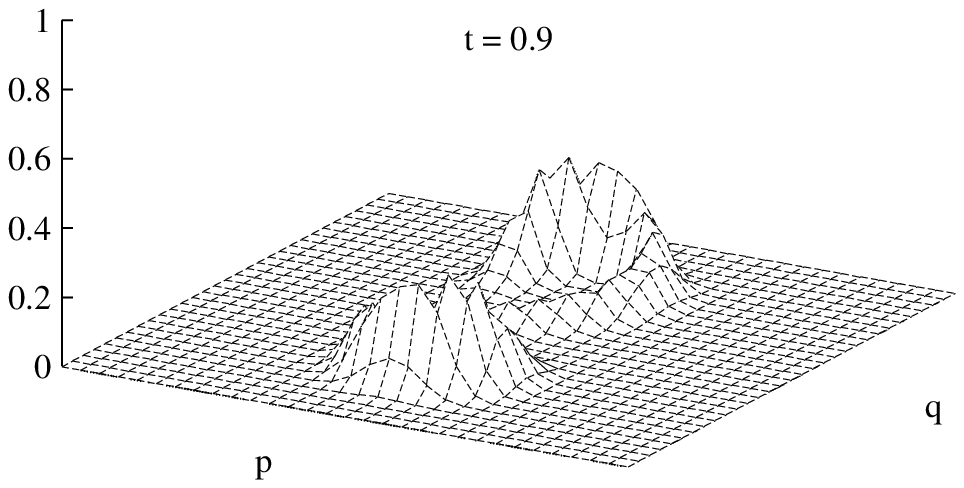,width=75mm}
}
\centerline{
\psfig{figure=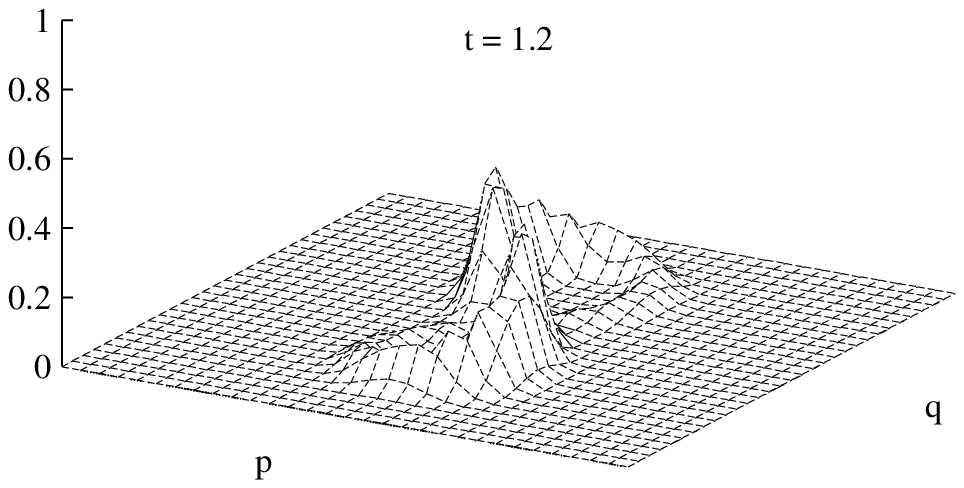,width=75mm}
\psfig{figure=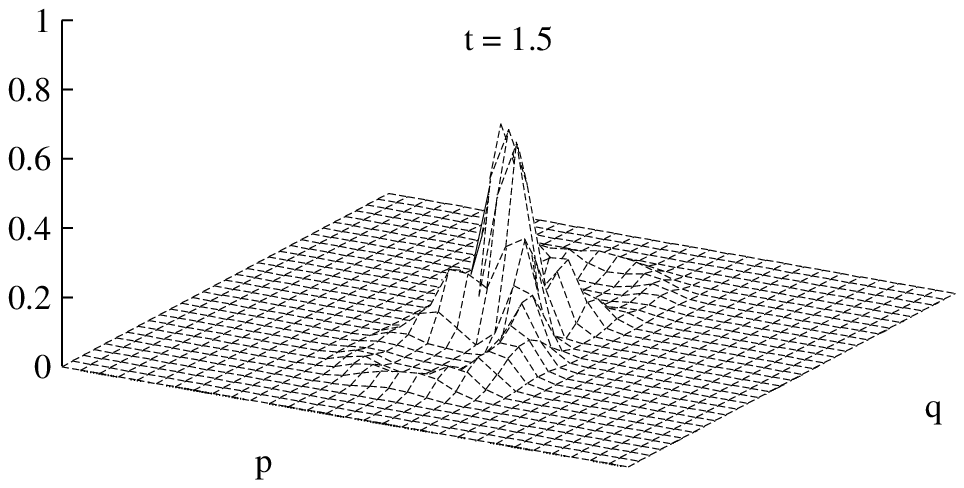,width=75mm}
}
\caption{\label{wp-ini}
  Evolution of a Gaussian wave packet initially located at the
  hyperbolic fixed point for $p=2$ and $r=0.01$. Husimi distribution
  (\ref{husdist}) for the wave function projected on the excitonic
  state $z=0$, $\phi=0$ in a surface plot.}
\end{figure}

\begin{figure}
\centerline{
\psfig{figure=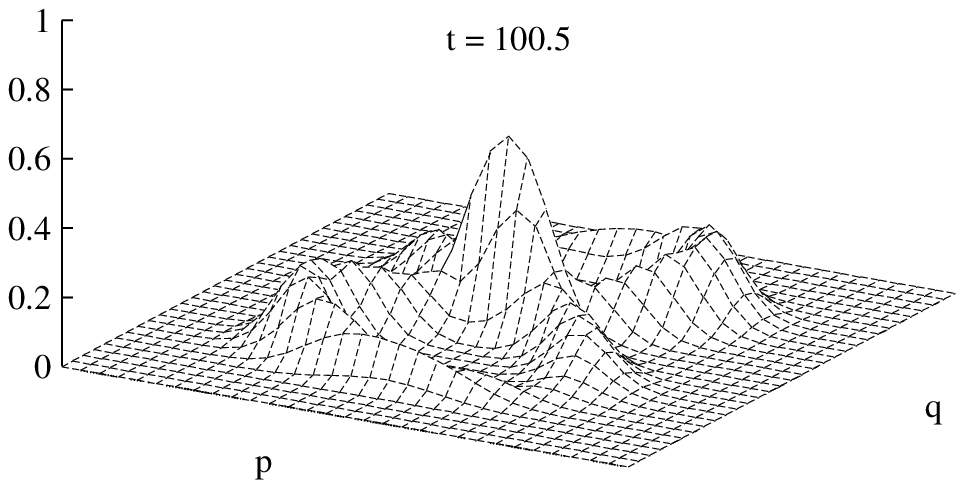,width=75mm}
\psfig{figure=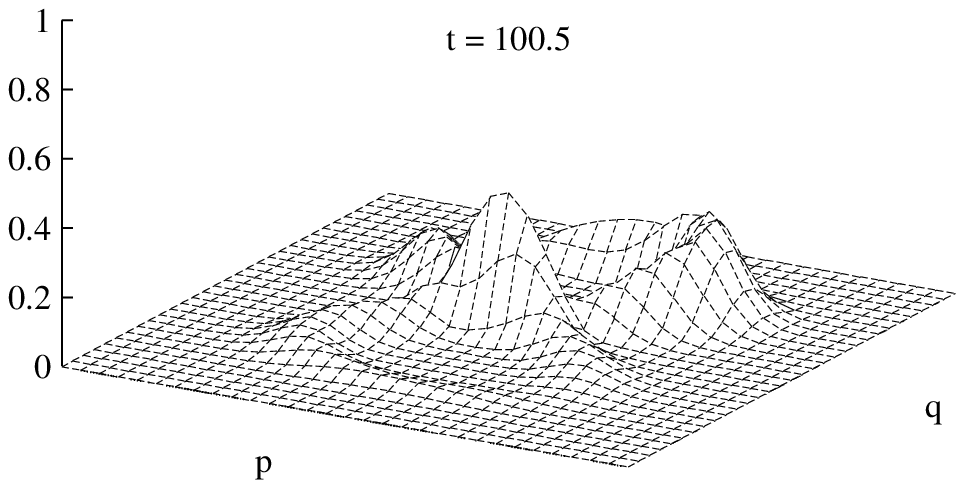,width=75mm}
}
\caption{\label{wp-lt}
  Form of a Gaussian wave packet initially located at the hyperbolic
  fixed point for $p=2$ and $r=0.01$ at large time. Husimi
  distribution of the wave function projected on the excitonic state
  $z=0$, $\phi=0$ (left) and $z=1$ (right) in a surface plot.}
\end{figure}

Of particular interest is the effect of the separatrix structure
characterising the mixed description above the bifurcation ($p>1$) on
the propagation of the oscillator wave packet. For a system with a
proper classical limit and a separatrix in the classical phase space
the correspondence to the quantum evolution was studied e.~g.\ in
\cite{Rei}. Similar to this we found that the presence of the
separatrix is clearly reflected in the wave packet dynamics when the
energy is fixed at $E_h$.  In the set of figs.\ \ref{wp-ini} the
evolution of a quantum state prepared initially right at the
hyperbolic fixed point $\bf h$ is presented.  The relevant system
parameters are $p=2$ and $r=0.01$ and the Husimi distribution
(\ref{husdist}) for the projection onto the excitonic state $z=0$,
$\phi=0$ is displayed. It is seen how the oscillator wave packet
spreads along the unstable direction of the separatrix structure. The
asymmetric distortion of the wave packet in the beginning of the
propagation, when the support of the Husimi distribution is given by
the unstable direction of the separatrix, is remarkable.  For long
times the wave packet covers the separatrix structure more uniformly
(see fig.\ \ref{wp-lt}(a)).

\begin{figure}
\centerline{
\psfig{figure=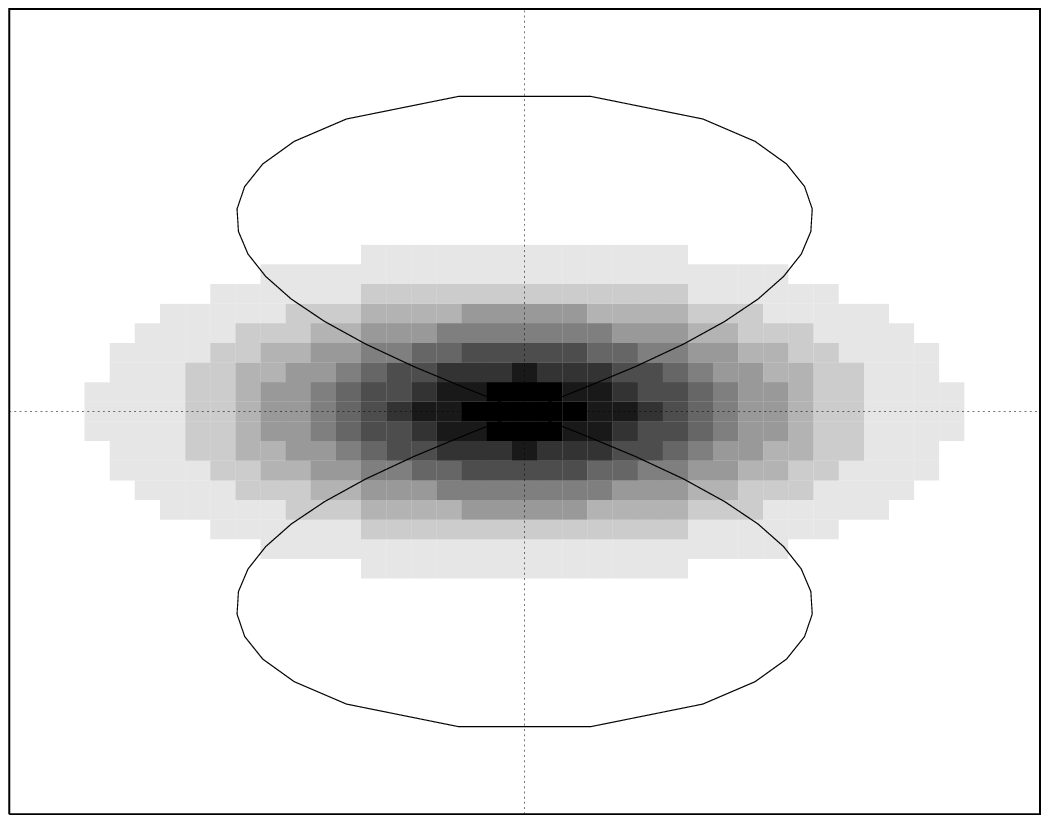,height=40mm,angle=90}
\psfig{figure=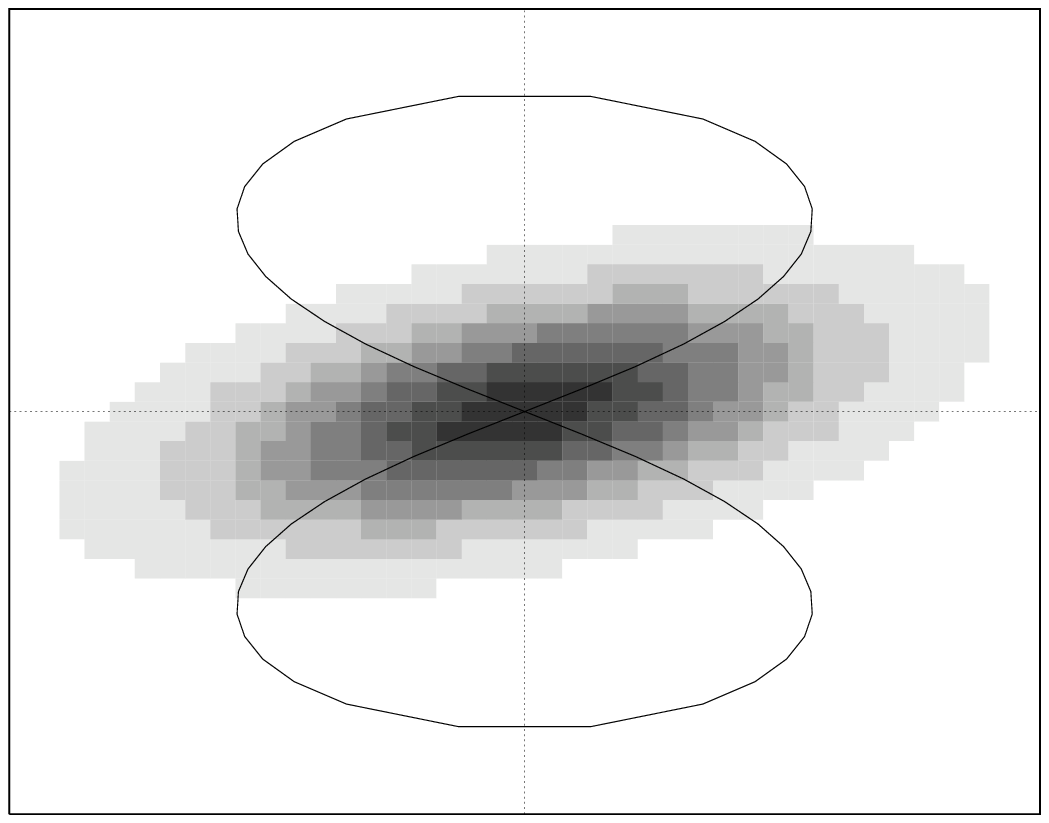,height=40mm,angle=90}
\psfig{figure=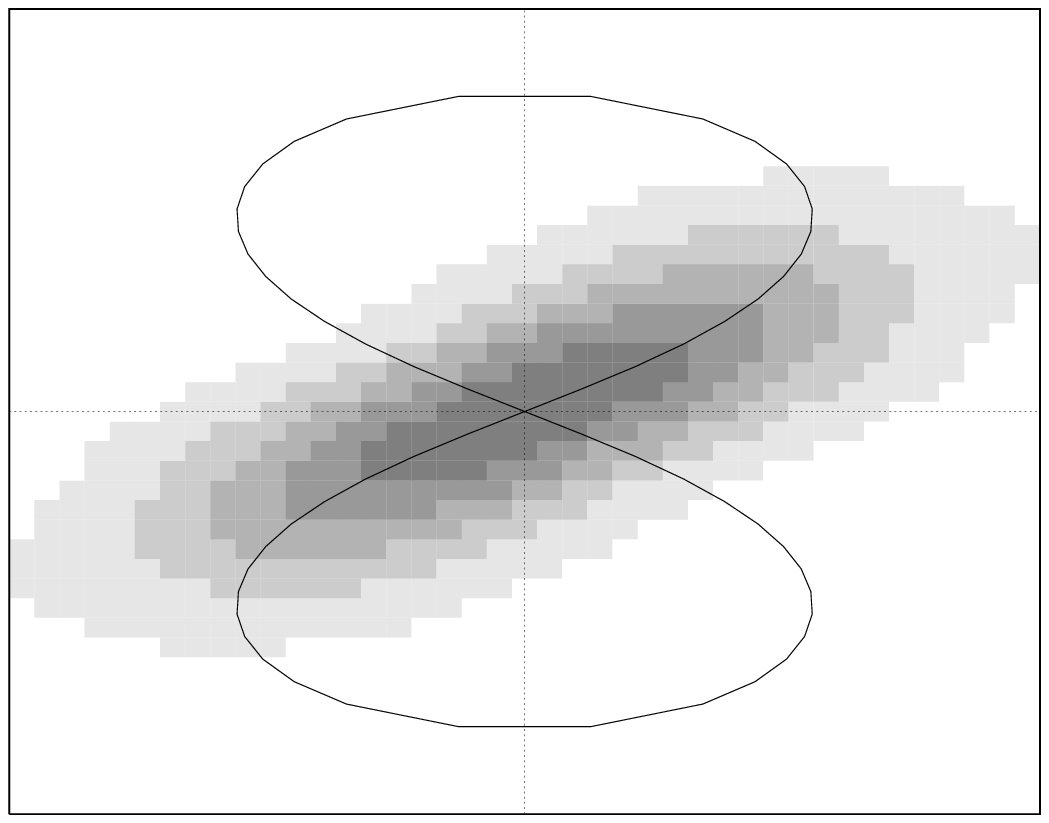,height=40mm,angle=90}
\psfig{figure=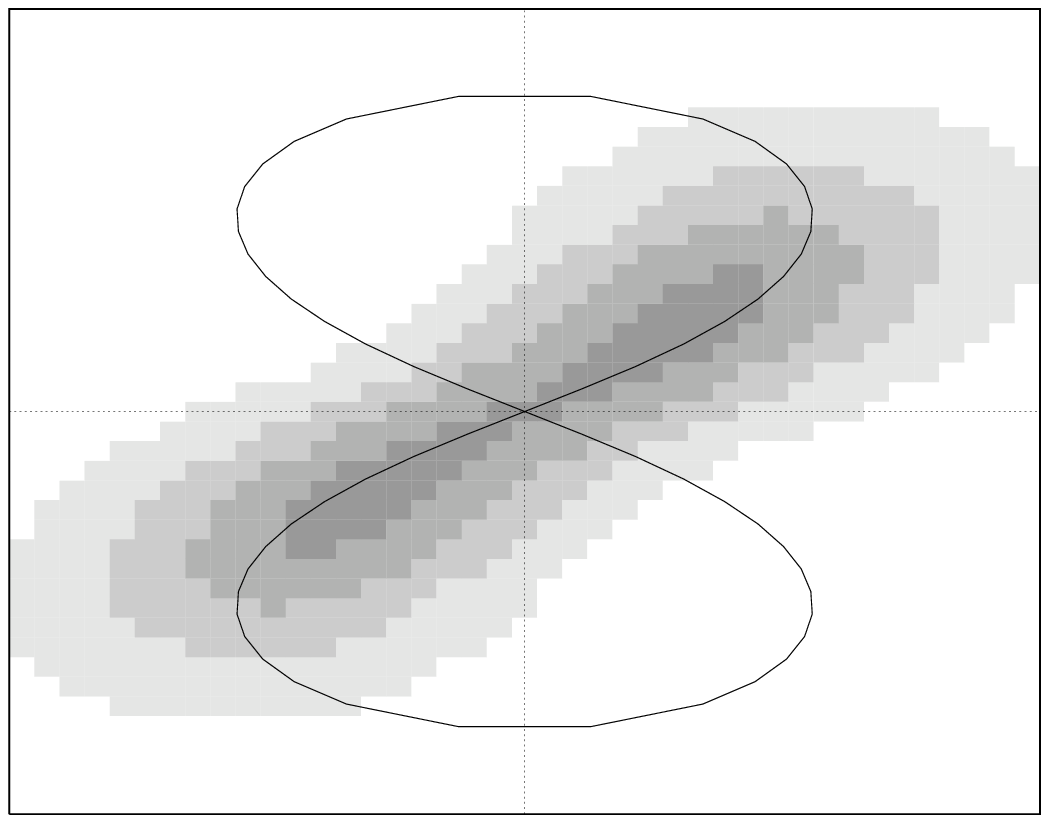,height=40mm,angle=90}
}                         
\centerline{              
\psfig{figure=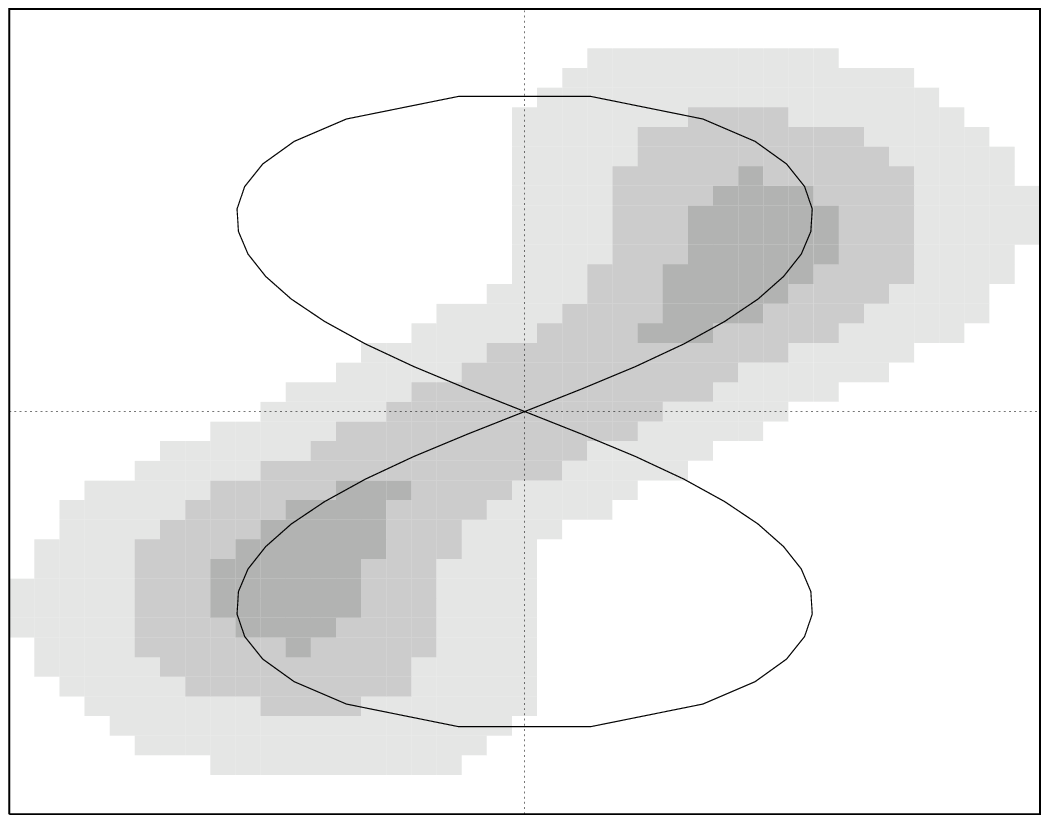,height=40mm,angle=90}
\psfig{figure=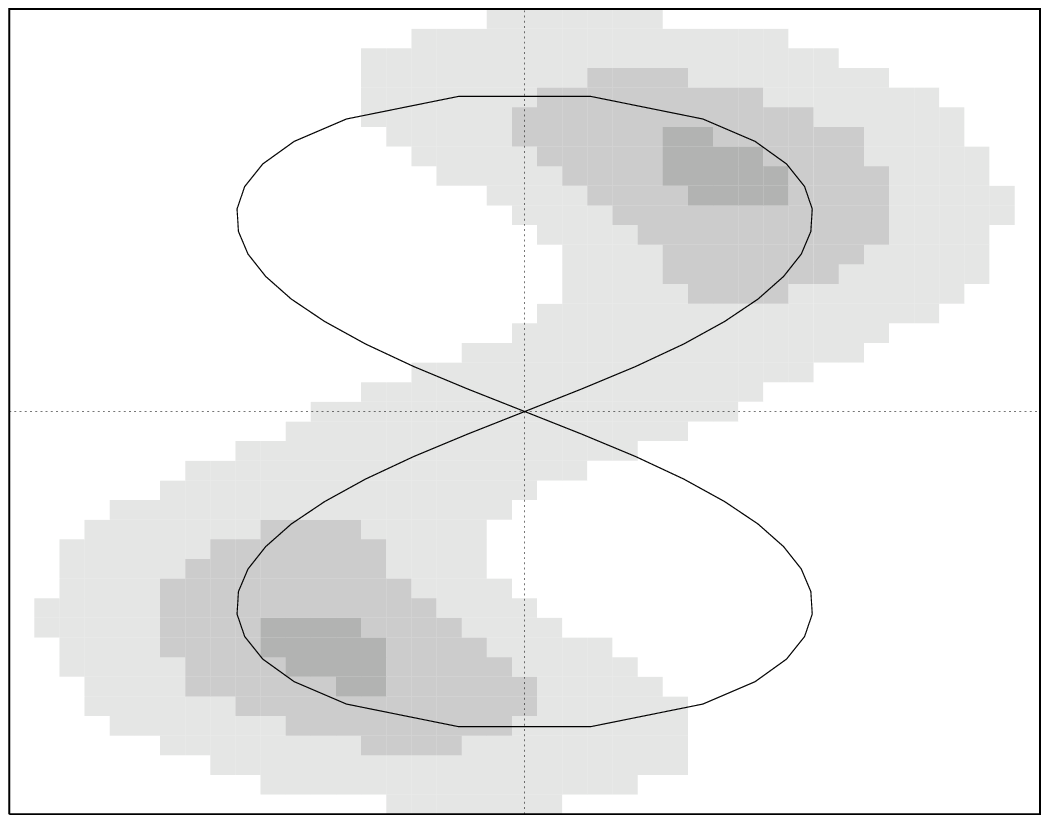,height=40mm,angle=90}
\psfig{figure=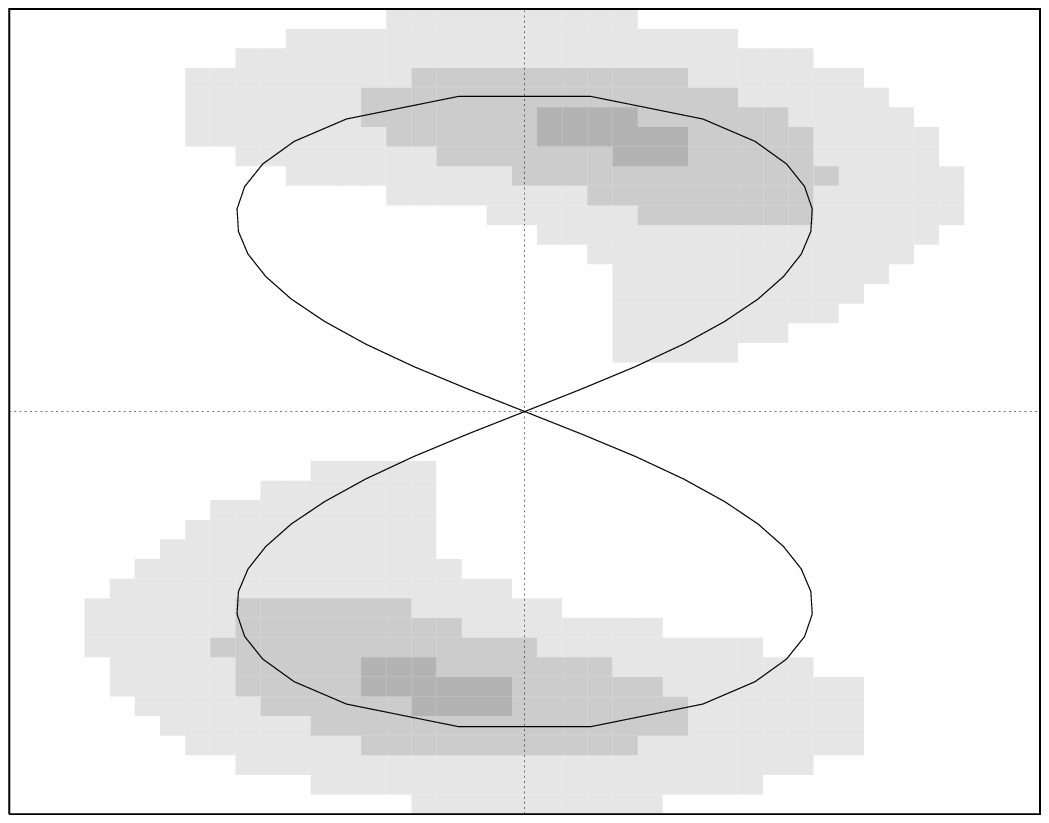,height=40mm,angle=90}
\psfig{figure=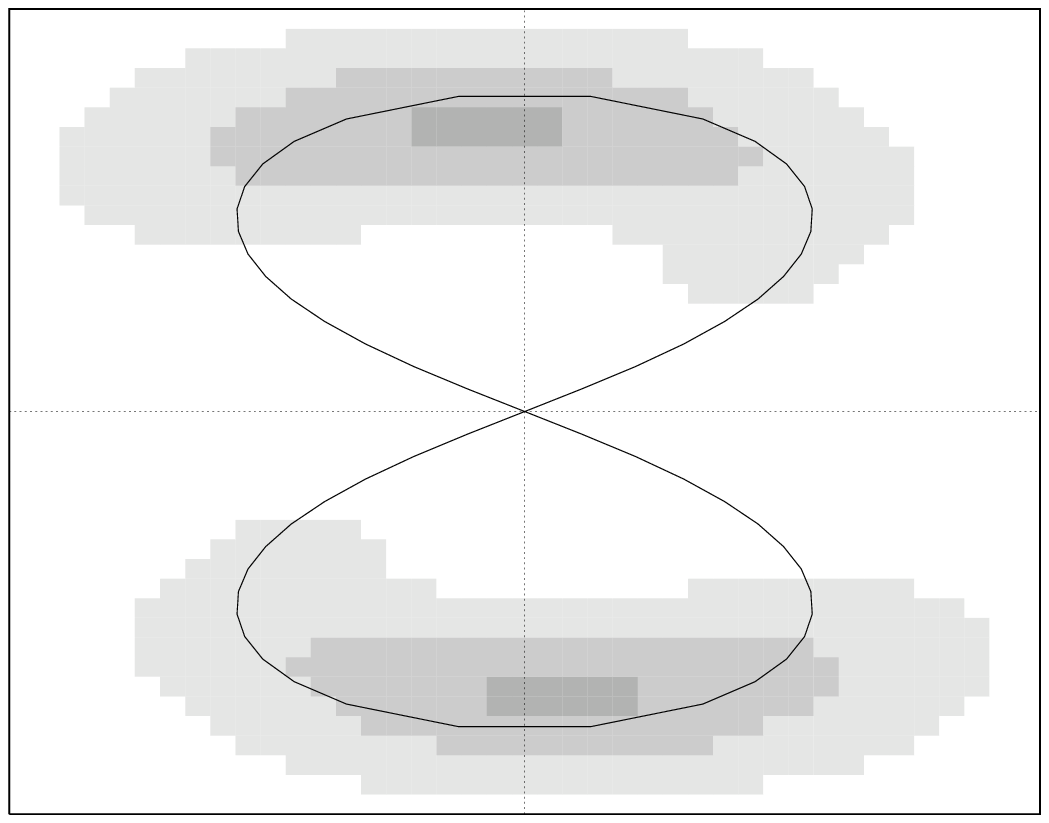,height=40mm,angle=90}
}                         
\centerline{              
\psfig{figure=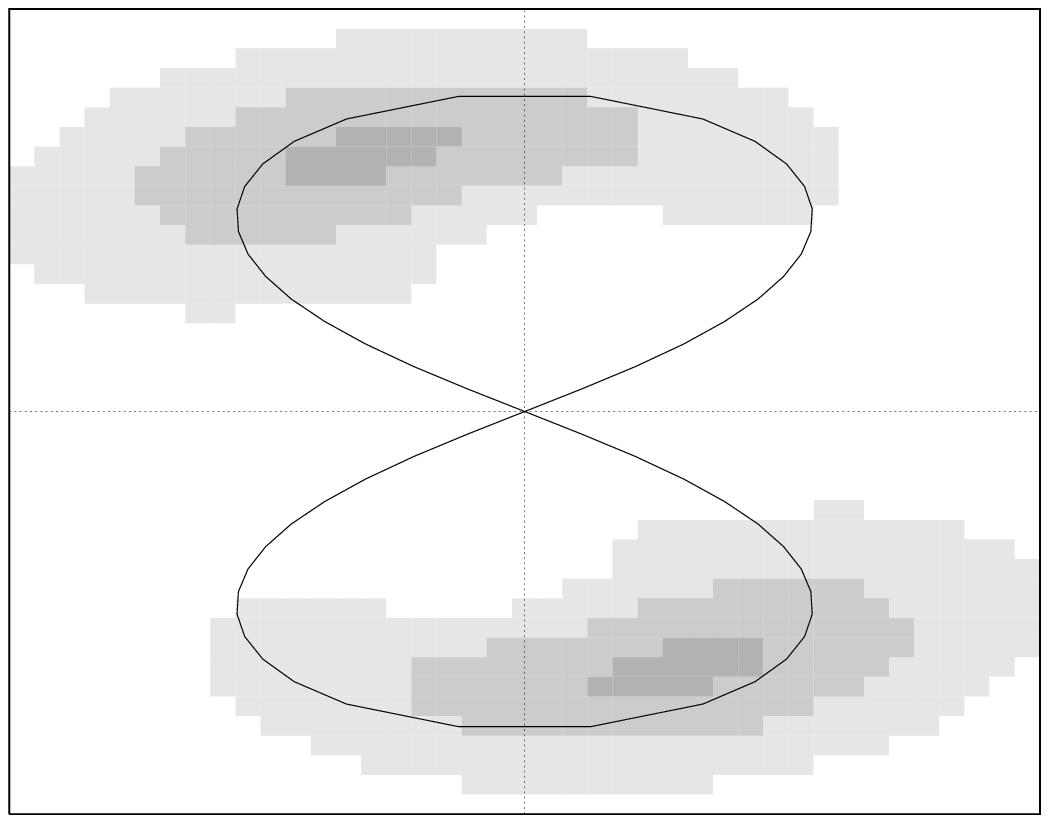,height=40mm,angle=90}
\psfig{figure=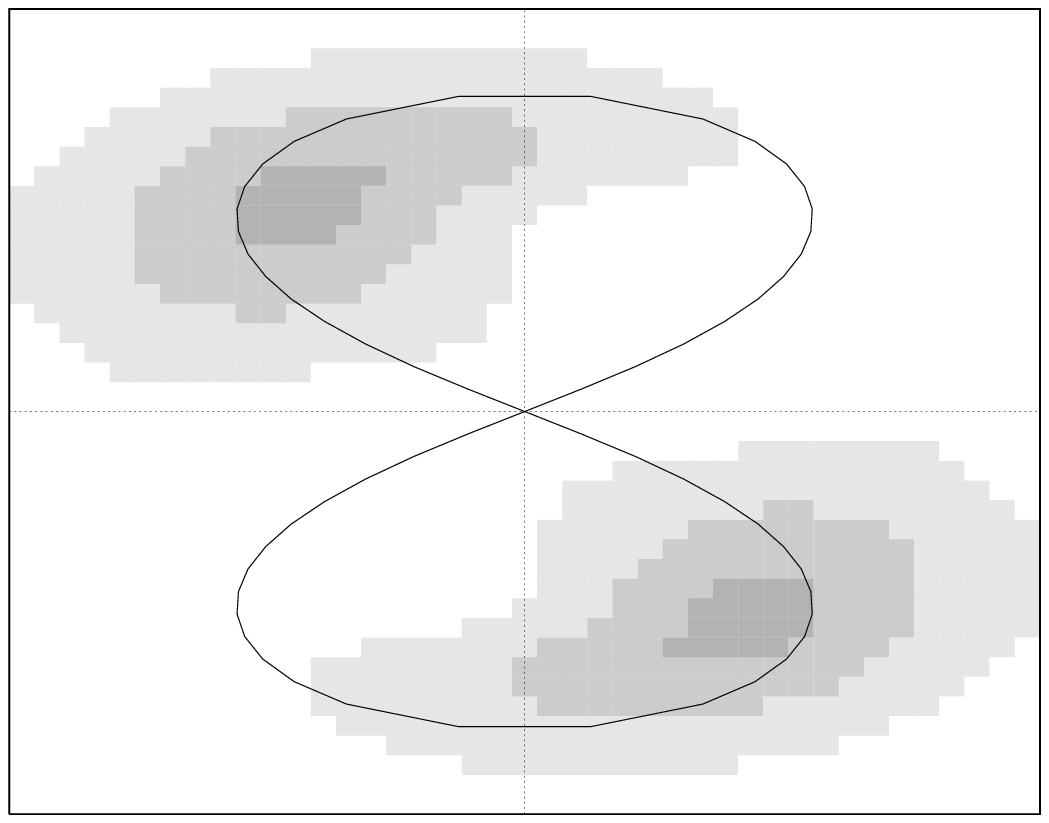,height=40mm,angle=90}
\psfig{figure=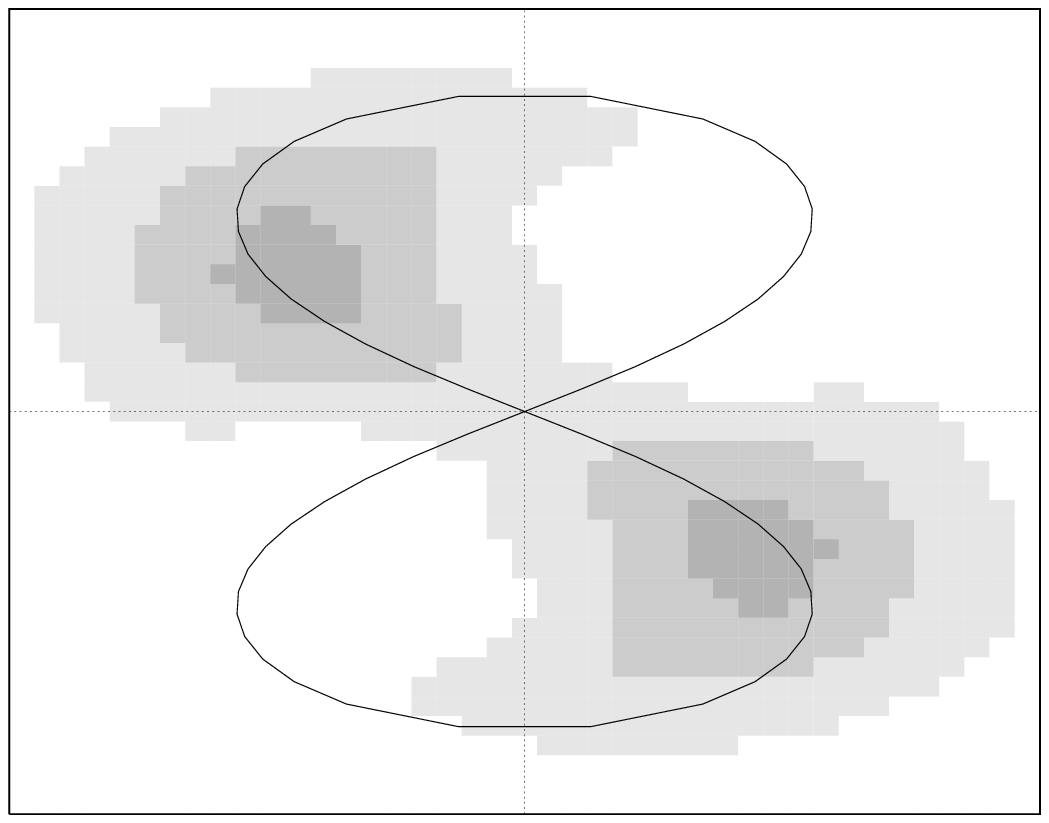,height=40mm,angle=90}
\psfig{figure=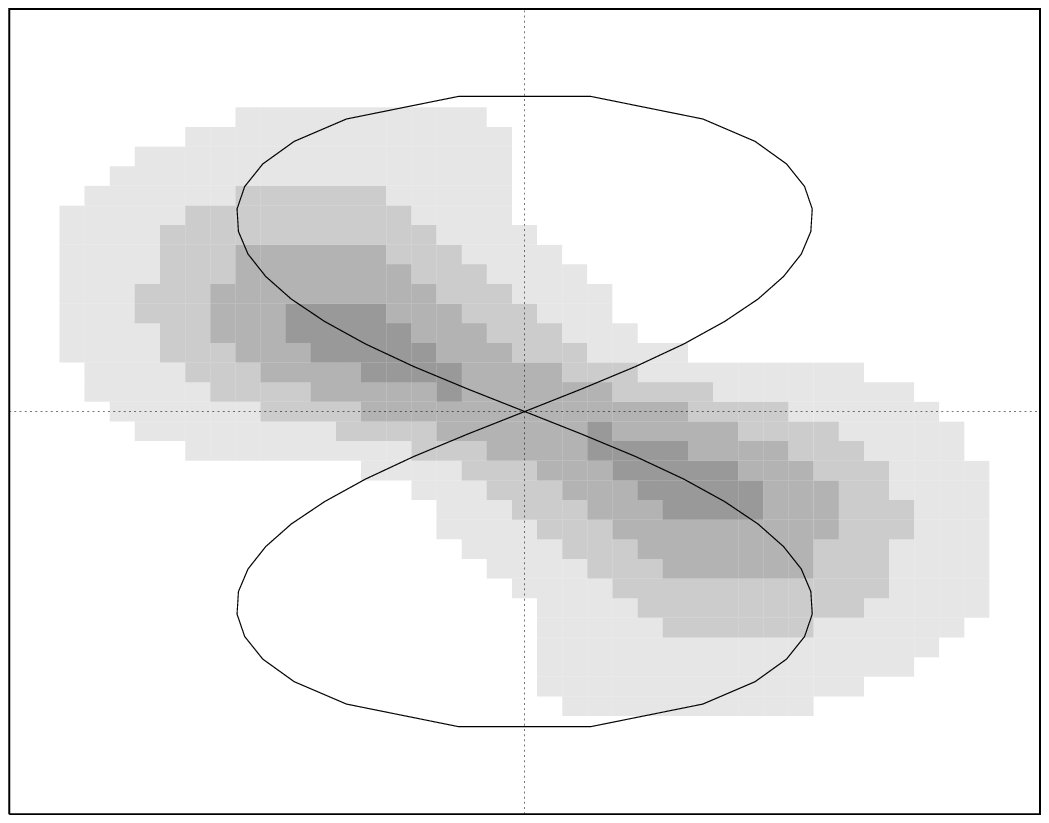,height=40mm,angle=90}
}                         
\centerline{              
\psfig{figure=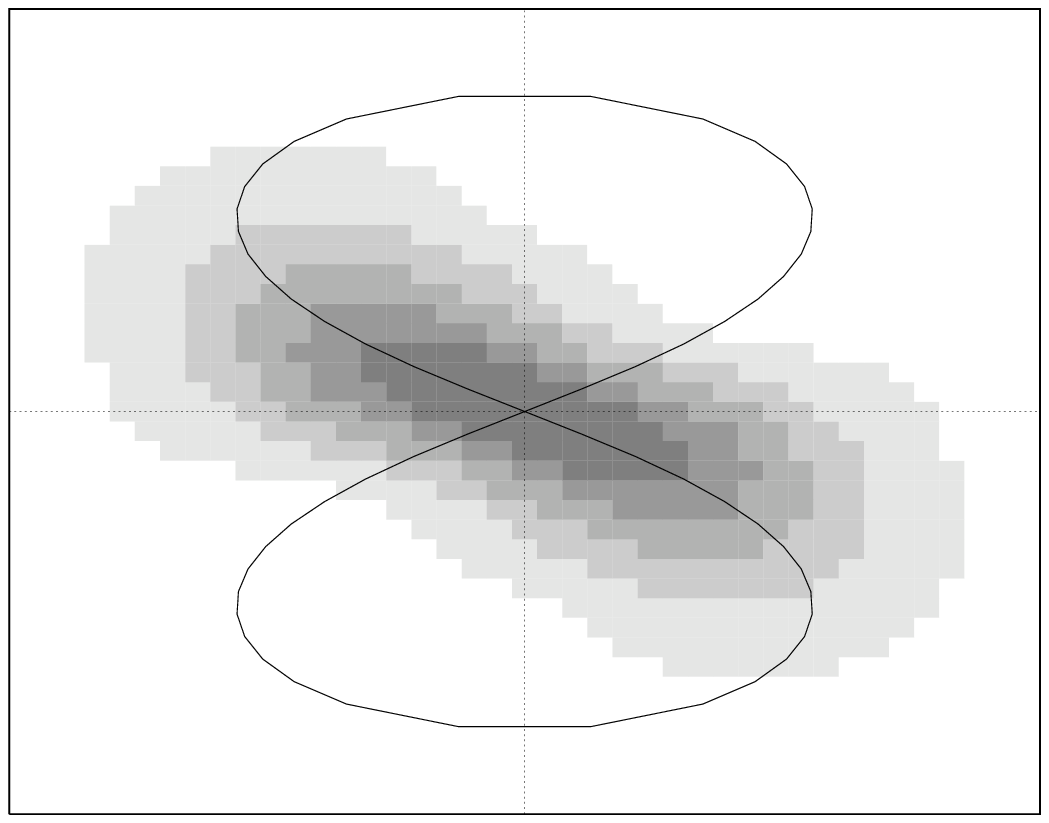,height=40mm,angle=90}
\psfig{figure=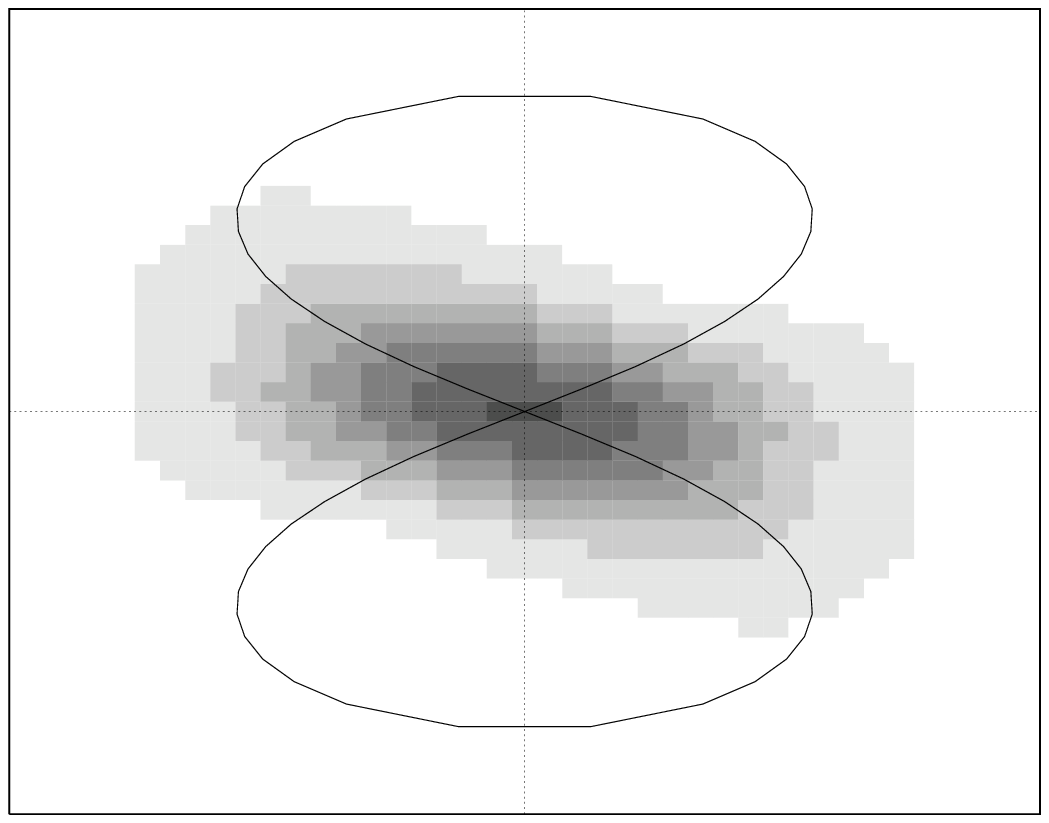,height=40mm,angle=90}
\psfig{figure=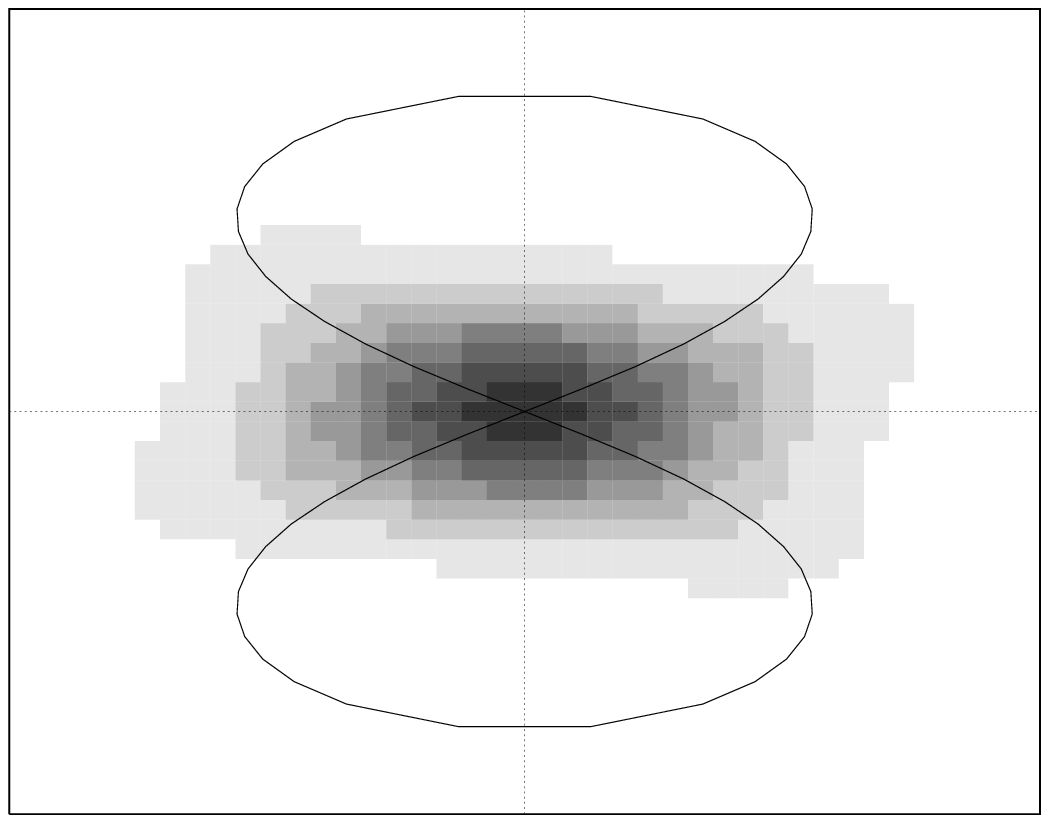,height=40mm,angle=90}
\psfig{figure=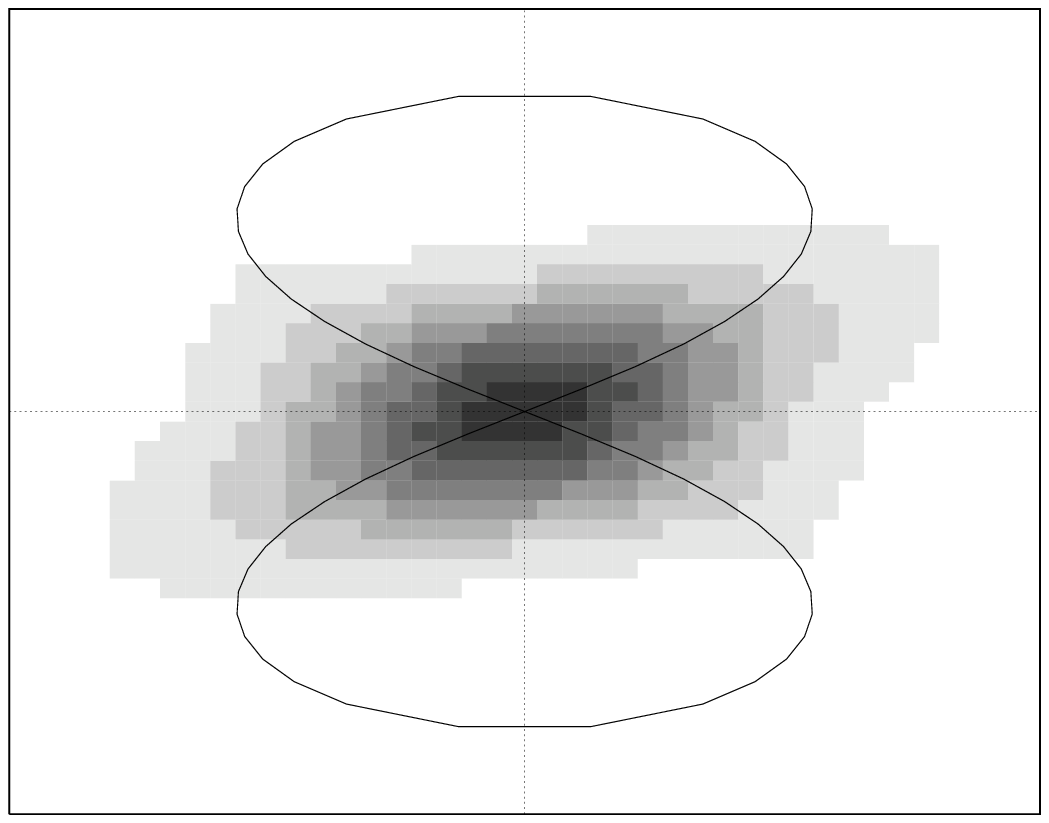,height=40mm,angle=90}
}
\caption{\label{hus-hp} Evolution of a quantum state prepared at $t=0$
  as Gaussian wave packet on the hyperbolic fixed point of the mixed
  quantum-classical dynamics ($p=2$, $r=0.1$, $\delta t=0.1$). The
  state is represented by the corresponding Husimi distribution after
  projection on the excitonic state $z=0$, $\phi=0$. The value of the
  Husimi distribution is encoded by the color using a linear scale
  with white corresponding to zero and black to one.  The solid line is the
  isoenergy line of the lower adiabatic potential at the energy of the
  wave packet ($E=-0.5$), i.~e.\ the separatrix associated with the
  hyperbolic fixed point in the adiabatic approximation.  }
\end{figure}

In the set of figs. \ref{hus-hp} contour plots for an analogous wave
packet propagation started at the hyperbolic point but for a larger
adiabatic parameter $r=0.1$ are presented.  The propagation along the
separatrix structure, which is now indicated by a full line, is again
evident.  This indicates that the well known classical-quantum
correspondence in the case of regular dynamics, namely that the
quantum distribution corresponds to the orbit of the corresponding
classical system, can be extended to systems treated in a mixed
quantum-classical description.  A more detailed comparison of the
results for $r=0.01$ (fig.\ \ref{wp-ini}) and $r=0.1$ (fig.\ 
\ref{hus-hp}) reveals as expected that the width of the wave packet
transversal to the underlying classical structure is reduced as the
system is closer to the adiabatic limit. We conclude that in the
adiabatic regime regular structures such as a separatrix in the
formally classical phase space of the mixed description can serve to
forecast qualitatively the evolution of a wave packet in the fully
quantized system.

In fig.\ \ref{wp-lt} we compare the Husimi distributions for one and the
same wave packet projected onto two different excitonic states in
order to reveal the quantum correlations between the excitonic and the
vibronic subsystems. In fig.\ \ref{wp-lt}(a) we chose $z=0$ and $\phi=0$
corresponding to equal site occupation probabilities whereas in fig.\ 
\ref{wp-lt}(b) the wave packet is projected onto $z=1$, i.~e.\ an
excitonic state completely localized at one of the dimer sites. It is
seen that for the case of an equal site occupation the oscillator
evolution proceeds along both branches of the separatrix structure
whereas for the one sided projection the oscillator is preferentially
located on the branch of separatrix corresponding adiabatically to the
occupied site.  This behavior reflects the property of the quantum
system to include coherently all the variants of motion of the mixed
quantum-classical system weighted with the corresponding probability
in analogy to the semiclassical propagator of a system with proper
classical limit, which is given as a sum over classical trajectories.

\begin{figure}
\centerline{\psfig{figure=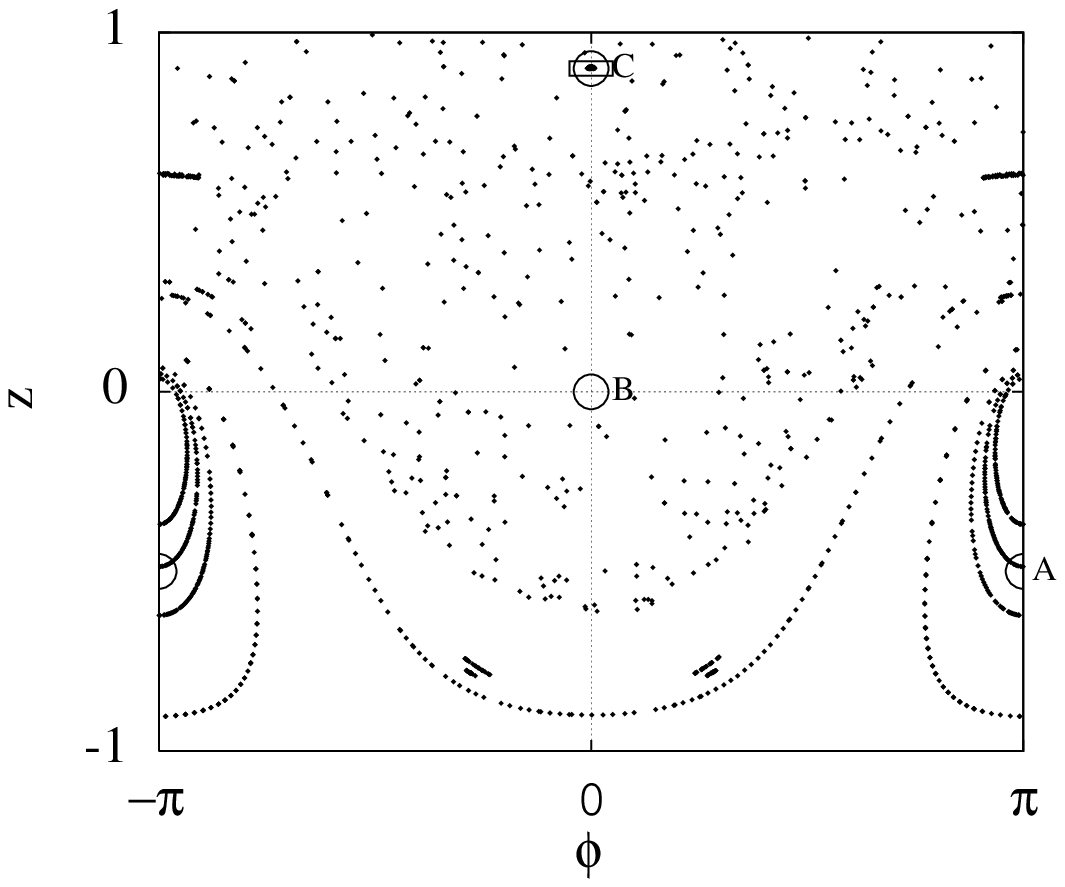,height=11cm}}
\vspace*{3mm}
\centerline{\hspace*{2mm}\psfig{figure=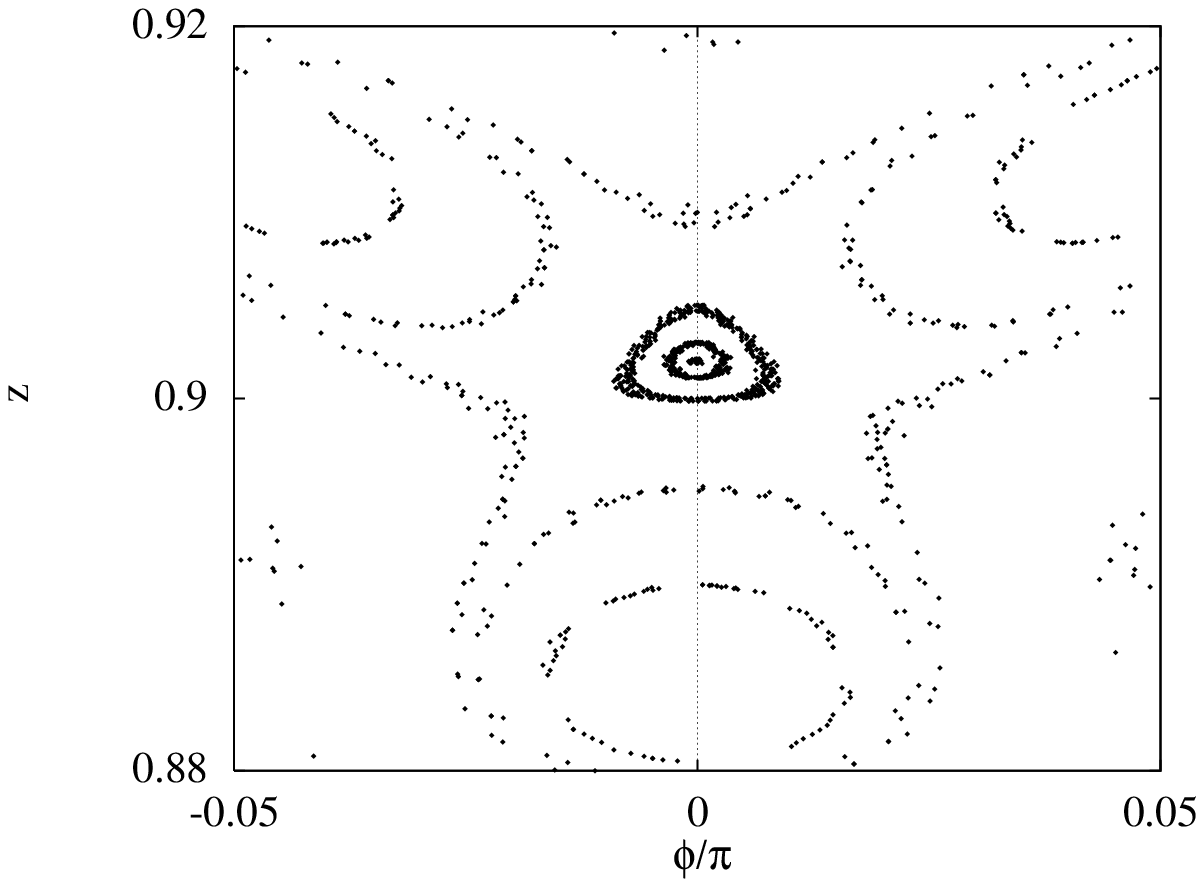,height=55mm}}
\caption{\label{icond}
Poincar\'e section $P = 0$, $dP/dt > 0$ for the symmetric dimer with
$r = 0.1$ and $p = 2$ at $E = +0.5$. The circles mark the initial
states, for which the time evolution will be displayed in the
following figures. A regular island embedded into the chaotic sea has
been enlarged in the lower part of the figure.  }
\end{figure}

\begin{figure}
\centerline{\psfig{figure=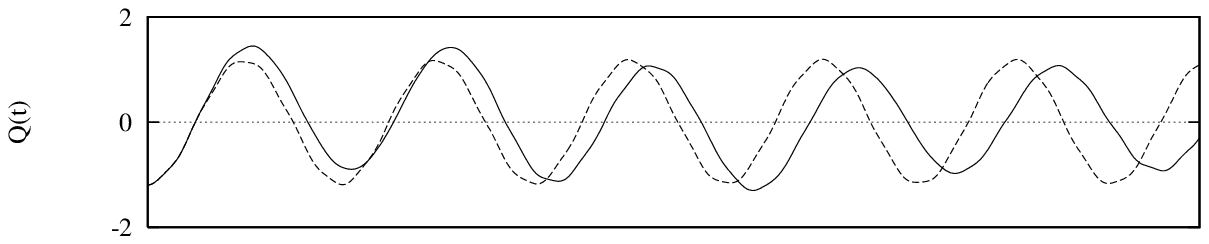,width=13cm}}
\centerline{\psfig{figure=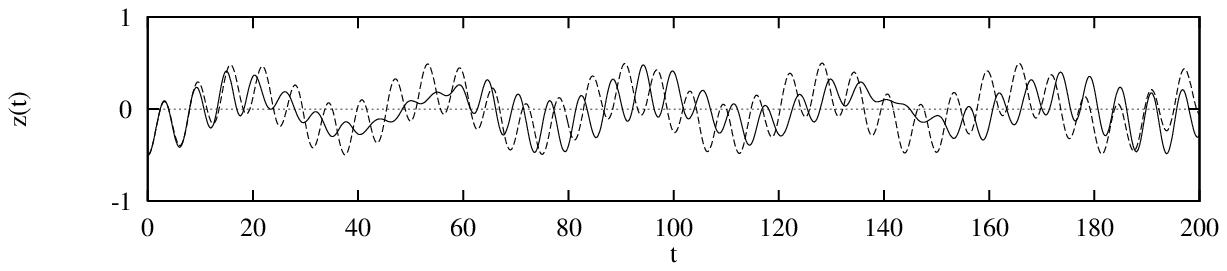,width=13cm}}
\centerline{\psfig{figure=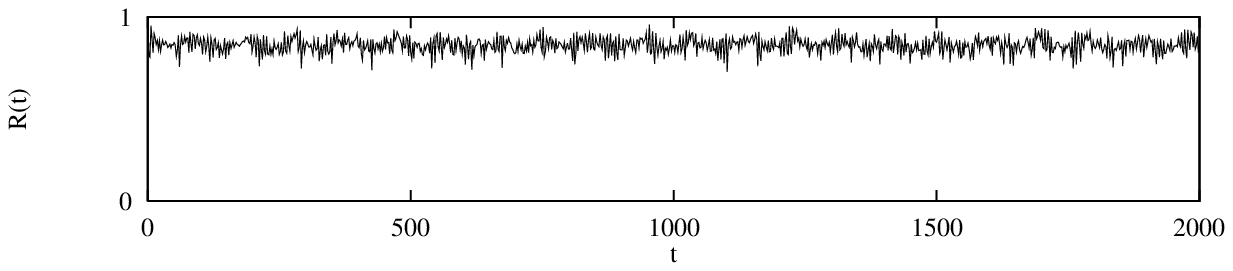,width=13cm}}
\caption{\label{tdepA}
  Time evolution of the initially factorized coherent state
  corresponding to {\bf A} in fig.~\ref{icond} ($\phi(t=0)=\pi$,
  $z(0)=-0.5$, $Q(0)=-1.2$, $E=0.55$). In the upper parts the quantum
  expectation values of $Q$ and $\sigma_z$ (full lines) are compared to
  the corresponding quantities of the mixed quantum-classical
  description (dashed lines). In the lower part the quantum Bloch
  radius $R(t)$ (\ref{qbr}) is displayed using a larger time scale.}
\end{figure}

\begin{figure}
\centerline{\psfig{figure=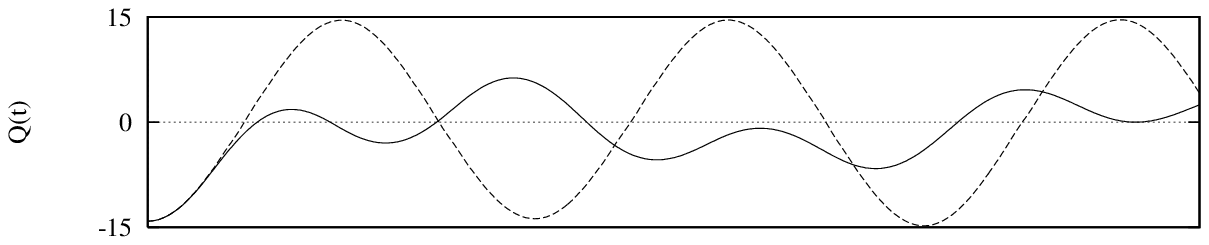,width=13cm}}
\centerline{\psfig{figure=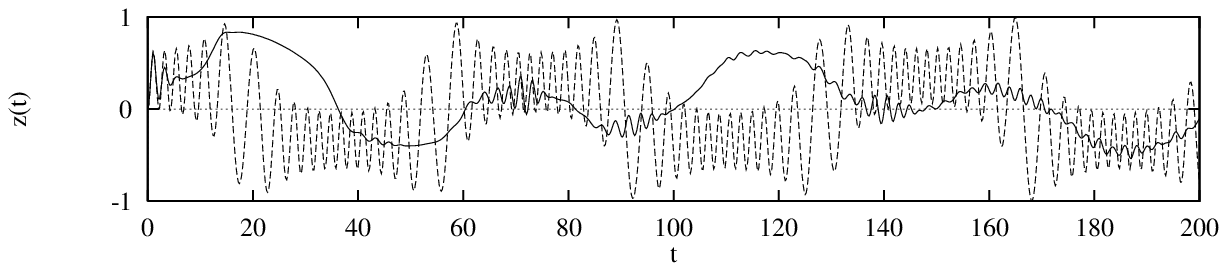,width=13cm}}
\centerline{\psfig{figure=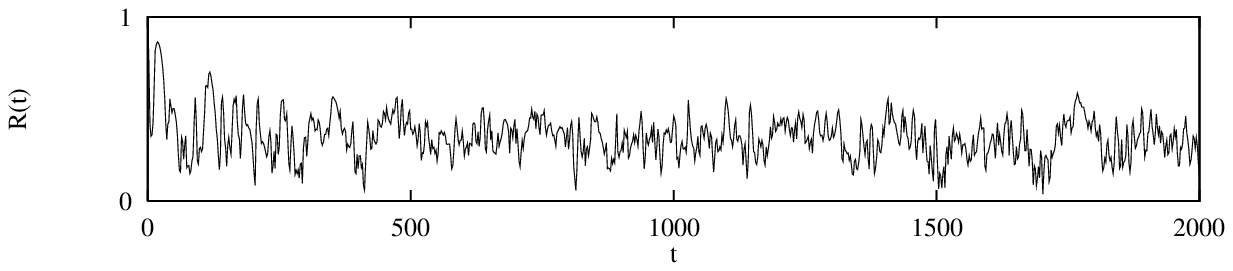,width=13cm}}
\caption{\label{tdepB}
  Time evolution of the initially factorized coherent state
  corresponding to {\bf B} in fig.~\ref{icond} ($\varphi(\tau=0)=0$,
  $z(0)=0$, $Q(0)=-14.1$, $E=0.55$).  See fig.\ \ref{tdepA} for
  details.  }
\end{figure}

\begin{figure}
\centerline{\psfig{figure=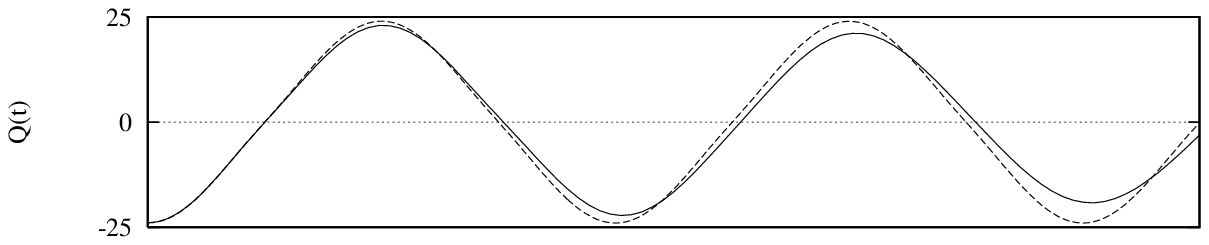,width=13cm}}
\centerline{\psfig{figure=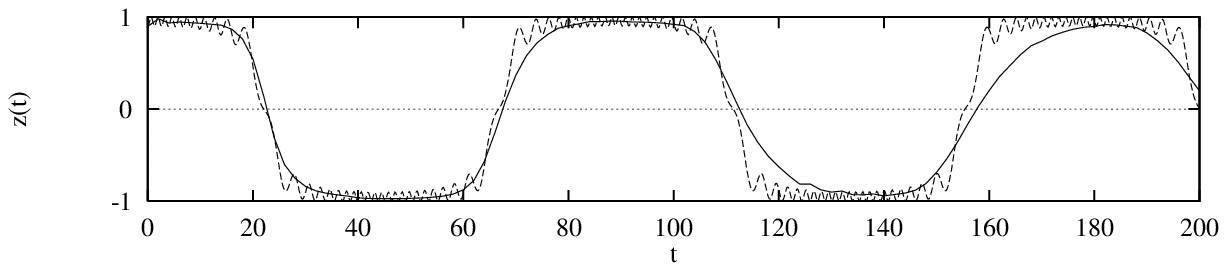,width=13cm}}
\centerline{\psfig{figure=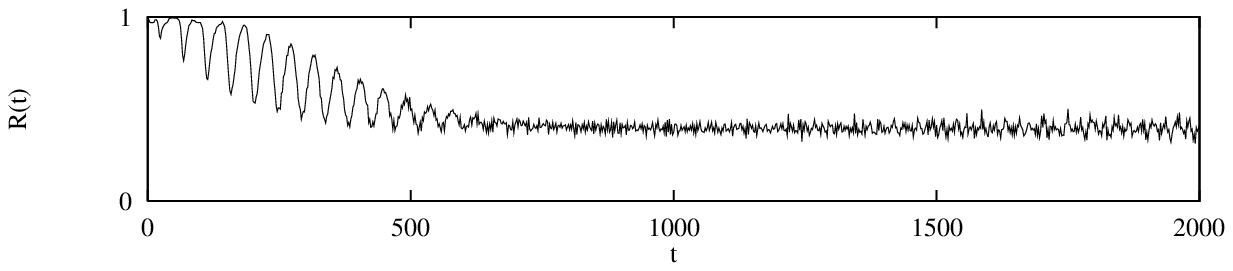,width=13cm}}
\caption{\label{tdepC}
  Time evolution of the initially factorized coherent state
  corresponding to {\bf C} in fig.~\ref{icond} ($\varphi(\tau=0)=0$,
  $z(0)=0.9$, $Q(0)=-24.0$, $E=0.55$).  See fig.\ \ref{tdepA} for
  details.  }
\end{figure}

Finally we address the problem of how the qualitative differences
between the regular and the chaotic dynamics of the system in the
mixed description are reflected in the evolution of the fully
quantized system, i.~e.\ whether there are signatures of the dynamic
chaos in the mixed description in the time dependent state vector of
the fully quantized system. For simple systems quantized in one step
and chaotic in the classical limit the differences in the quantum
evolution between initial conditions selected in the classical regular
and chaotic parts of the phase space of the system are well known: If
e.~g.\ the initial conditions of the quantum system are selected in
the regular part of the classical phase space the time dependence of
the appropriately chosen quantum expectation values follow ("shadow")
the corresponding classical values over a substantial amount of time,
whereas for initial conditions chosen in the chaotic part of the
classical phase space these dependences start to deviate from each
other almost immediately (see e.\ g.\ \cite{BBH}). In order to
investigate this connection in our case we have selected different
initial conditions in the regular and chaotic parts of the Bloch
sphere of the system and compared the evolution in the mixed
description with that of expectation values obtained from the fully
quantized system. In fig.\ \ref{icond} the location of three different
initial conditions on the Bloch sphere of the excitonic subsystem
belonging to the main regular (A) and chaotic (B) regions of the
dynamics in the mixed description, as well as a small regular island
(C) embedded in a large chaotic surrounding are shown.  For a
comparison of the dynamics in the mixed and fully quantized
descriptions for these cases we selected the variables $Q(t)$ and
$z(t)$ displayed in the upper parts in the set of figs.\ 
\ref{tdepA}-\ref{tdepC}.

We first compare the dynamics for initial conditions located in the
main regular (antibonding) and main chaotic (bonding) regions. Since
the initial state of the fully quantized system is chosen as a product
state with factorizing expectation values for which the decoupling
implicit in the derivation of (\ref{eommix}) is justified, there is
always an interval at the beginning of the time evolution where the
mixed description follows closely the quantum data.  Then, however,
there is indeed a striking difference between initial conditions
selected in the regular and the chaotic parts of the phase space of
the mixed system: For initial conditions in the regular part ({\bf A},
fig.\ \ref{tdepA}) the quantum expectation value $Q(t)$ follows
closely the classical trajectory of the mixed description over several
periods and then, apart from a slowly growing phase shift, both
dependences keep a similar oscillatory form, whereas for an initial
condition in the chaotic part ({\bf B}, fig.\ \ref{tdepB}) the
corresponding curves are completely different and the deviation
between both starts already after a fourth of the oscillator period.
This confirms for our case the general behavior of classically chaotic
systems to produce a fast breakdown of the validity of quasiclassical
approximations when quantum effects become important. The comparison
for the occupation difference $z(t)$ of the excitonic sites is not so
direct, because the exciton constitutes the fast subsystem resulting
in rapid oscillations of $z(t)$ in the mixed description. However, for
the regular case we observe that the slowly changing mean value of
$z(t)$ obtained from the mixed description is related to the quantum
data, though amplitude and phase of the superimposed rapid
oscillations are different after a few periods of the excitonic
subsystem. In the chaotic case the breakdown of the mixed description
for shorter times is evident and there is no correspondence for the
mean values.

The gradual development of quantum correlations between both
subsystems, which are absent in the initially factorized state, can be
quantified by calculating the effective Bloch radius 
\begin{equation}\label{qbr}
R(t)=\sqrt{x(t)^2+y(t)^2+z(t)^2}
\end{equation}
of the excitonic subsystem using the time dependent expectation values
of $\sigma_x$, $\sigma_y$ and $\sigma_z$. Note that the reduced density
matrix $\rho(t)$ for the excitonic subsystem, obtained from the full
density matrix by taking the trace over the oscillator states, is
related to $R(t)$ via ${\rm Tr}\, \rho(t)^2=(1/2)(1 + R(t)^2)$.  For the
factorized and correspondingly uncorrelated initial quantum state the
value of the Bloch radius is $R=1$ and $R(t)$ will decrease in the
course of time according to the degree to which quantum correlations
lead to an entanglement between both subsystems.  In the lower parts
of the figs.\ \ref{tdepA}-\ref{tdepC} the dependence of the Bloch
radius on the time is displayed for a long time interval.  The
difference between the behavior for initial conditions chosen in the
regular and chaotic parts of the phase space of the mixed description
is remarkable: For initial conditions in the regular part of the phase
space after an initial drop $R(t)$ stabilizes at a value close to $1$,
whereas for the initial conditions in the chaotic part the descent is
much more pronounced and the long time value of $R(t)$ is much lower,
thus indicating stronger quantum correlations in the chaotic case. The
correlations between the subsystems are the reason for the breakdown
of the mixed description which implicitly contains the factorization
of expectation values. Therefore the smaller value of $R(t)$ observed
for the state prepared in the chaotic region confirms the faster
breakdown of the mixed description as compared to a regular initial
state. However, it is important to note that the striking difference
between the values of $R(t)$ is not restricted to this initial period
but extends to much longer times (which are on the other hand small
compared to the time for quantum recurrences). In this respect our
results indicate time dependent quantum signatures of chaos of the
mixed description which are beyond the well known different time
scales for the breakdown of quasiclassical approximations.

Finally we present the example for a quantum state prepared on a
regular island embedded into chaotic regions of the mixed
quantum-classical phase space ({\bf C}, fig.\ \ref{tdepC}). The
structure of the selected island is shown in the lower part of fig.\ 
\ref{icond}. For such a state the situation is specific due to the
spreading of the quantum state out of the regular island.  After some
initial time in which the quantum dynamics probes the regular region
of the the mixed dynamics the wave packet enters the region in which
the mixed dynamics is chaotic.  Correspondingly we find for an initial
time interval that the agreement between the mixed and the full
quantum description is as good as expected for regular dynamics
whereas for long times the quantum system shows the typical behavior
of a chaotic state. This is evident from the time dependence of the
Bloch radius which is displayed in the lower part of fig.\ \ref{tdepC}
on a sufficiently large time scale.

\pagebreak
\section{Conclusions}
1. We considered the nonlinear dynamical properties of a coupled
quasiparticle-oscillator system and demonstrated that the regular
structures of the mixed quantum-classical description such as the
fixed points and the presence of a separatrix are associated with the
corresponding adiabatic approximation, in which the nonadiabatic
couplings are switched off and integrable reference systems can be
defined. Comparing the evolution of quantum wave packets to the mixed
quantum-classical description we found that regular structures of the
mixed description can serve as a support for wave packet propagation
in the fully quantized system in the adiabatic regime. This should be
of interest for other systems to which a stepwise quantization must be
applied due to their more complex structure, e.~g.\ for the purpose of
forecasting the qualitative properties of propagating wave packets
using the mixed description as a reference system.

2. The nonadiabatic couplings, the inclusion of which is beyond the
adiabatic approximation, are identified as the source of dynamical
chaos observed in the mixed quantum-classical description. This
suggests that nonadiabatic couplings can be a general source of
nonintegrability and chaos also in other systems treated along a
stepwise quantization. Signatures of this type of chaos can then be
expected on the fully quantized level of description similar to what
we found for the coupled quasiparticle-oscillator system. In
particular, the breakdown of the mixed description is enhanced for
states prepared in a chaotic region of the phase space and the long
time evolution of these states is characterized by much stronger
quantum correlations between the subsystems.

3. Our results are related to the general question of how the idea of
the Born-Oppenheimer approach to analyze complex systems by a stepwise
quantization can be extended to a dynamical description. A more
systematic investigation of this question using other model systems is
certainly of interest in view of the widespread use of this approach.

\section{Acknowledgement}
Financial support from the Deutsche Forschungsgemeinschaft (DFG) 
is gratefully acknowledged.

\pagebreak

\pagebreak
\pagebreak
\begin{figure}
\begin{tabular}{cc}
\hspace*{0mm}
\psfig{figure=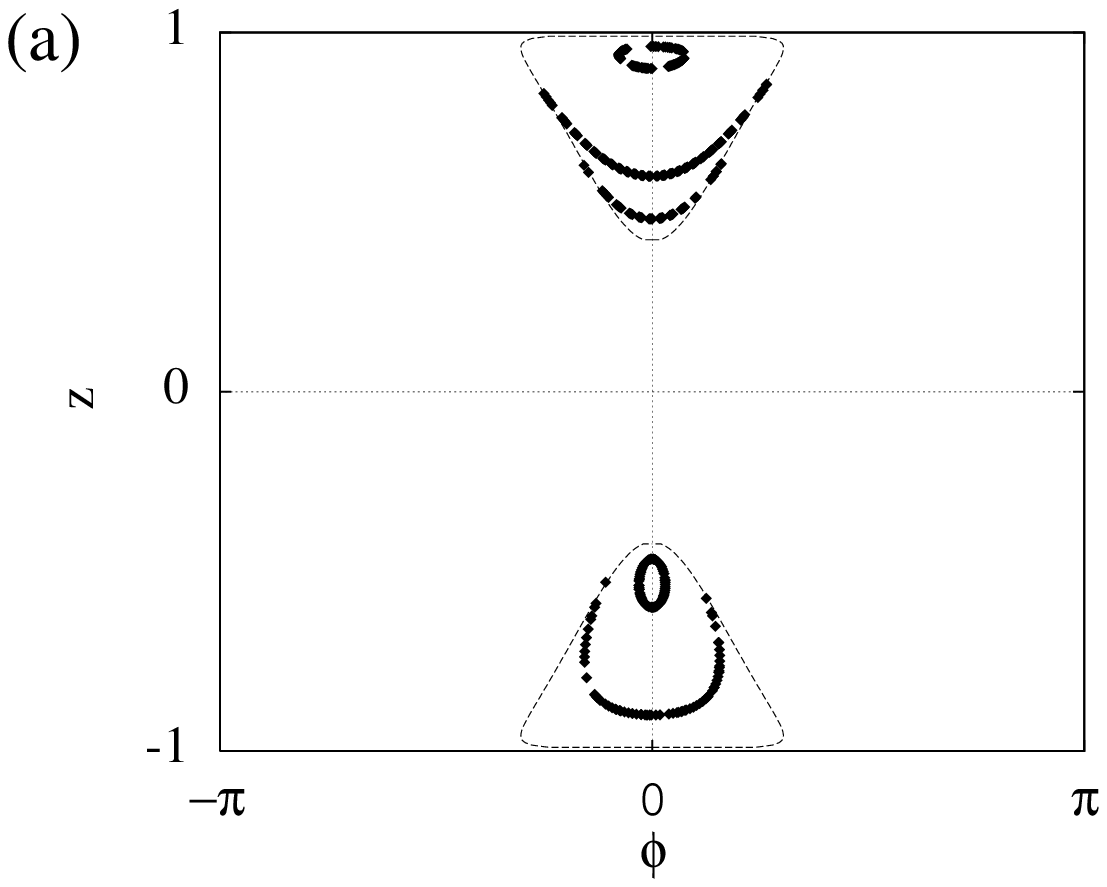,width=7cm,angle=0} &
\psfig{figure=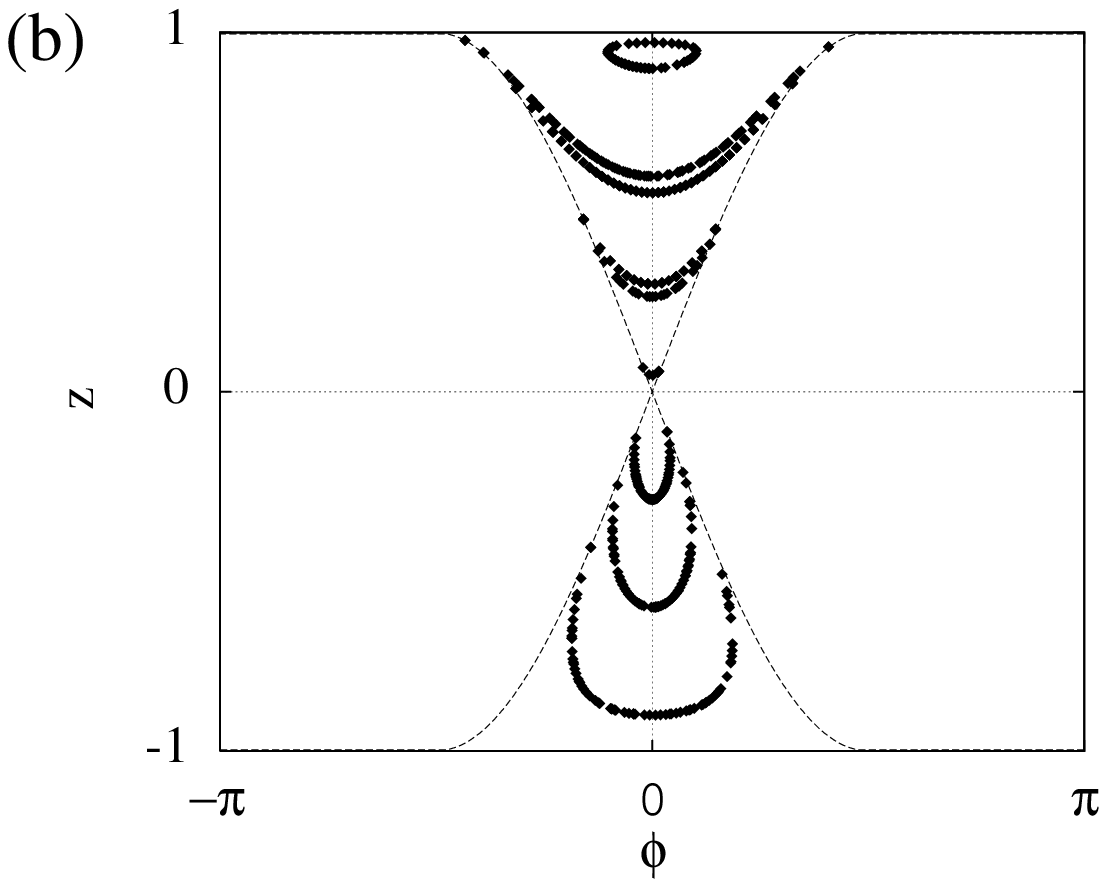,width=7cm,angle=0} \\
\hspace*{0mm}
\psfig{figure=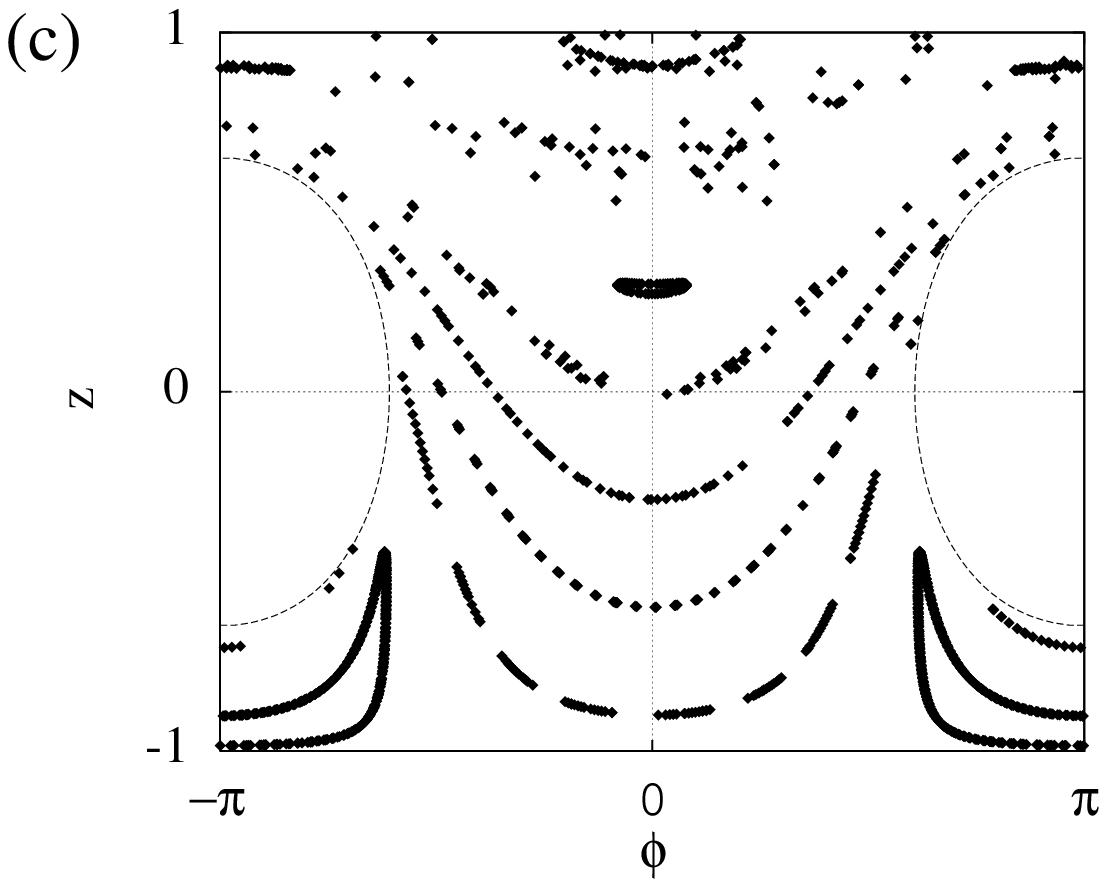,width=7cm,angle=0} &
\psfig{figure=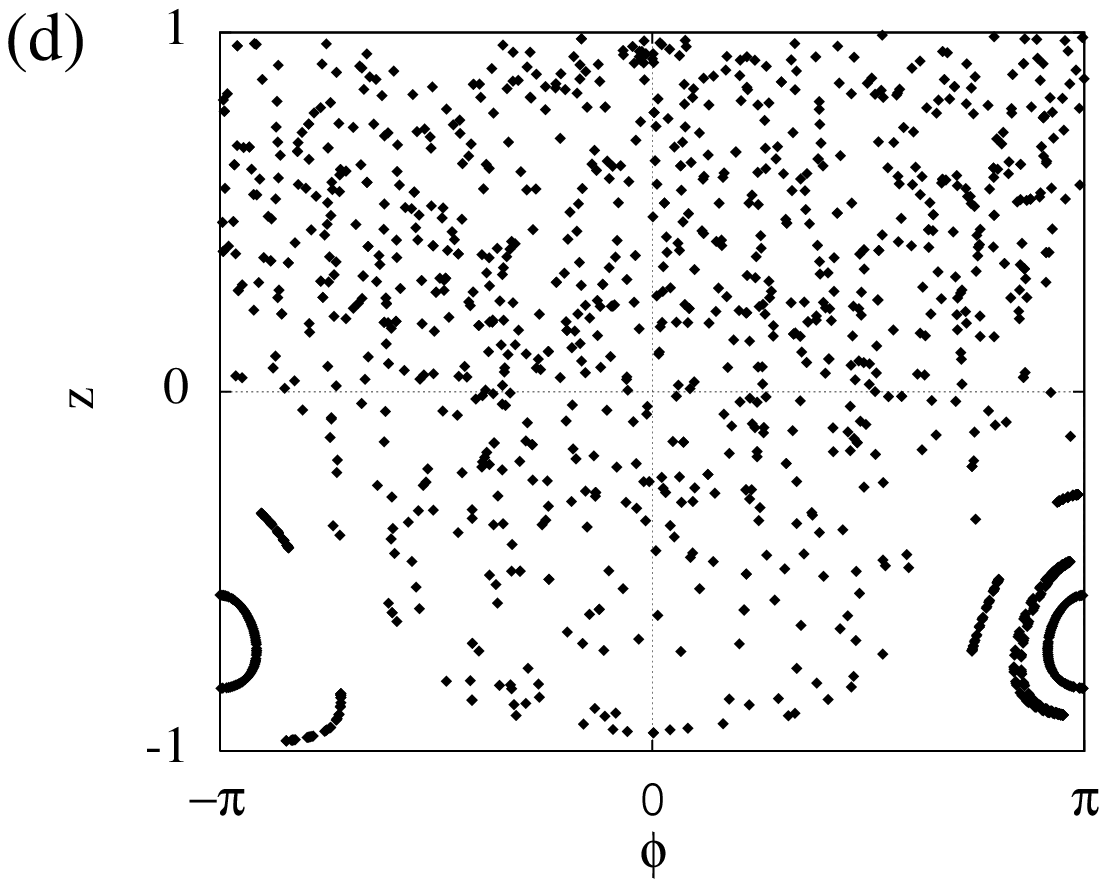,width=7cm,angle=0}
\end{tabular}
\caption{\label{03}
  Poincar\'e section in the excitonic variables for $r = 0.1$, $p =
  2.0$ (above bifurcation) and different energies: (a) $E = - 0.54$
  (b) $E = -0.5$ (c) $E = +0.5$ (d) $E = 0.83$ (see also fig.\ 
  \ref{icond}). The surface of section is defined by the left turning
  point of the oscillator ($P = 0$, $dP/dt > 0$) and displayed using
  the coordinates $z$ and $\phi$ (note the periodicity of the
  abscissa).}
\end{figure}
\pagebreak
\pagebreak
\pagebreak
\pagebreak
\pagebreak
\pagebreak
\pagebreak

\end{document}